\documentclass[10pt,journal,compsoc]{IEEEtran}
\usepackage{amsmath}
\usepackage{multirow}
\usepackage{color}
\usepackage{float}
\usepackage{cite}
\usepackage{filecontents}
\usepackage[utf8]{inputenc}
\usepackage{lmodern}
\usepackage{csquotes}
\usepackage[bottom]{footmisc}
\usepackage[justification=centering]{caption}
\usepackage{blindtext}
\usepackage{wrapfig}
\usepackage{graphicx}
\DeclareGraphicsExtensions{.eps,.pdf,.gif,.png,.jpg}
\usepackage{amssymb}
\usepackage{amsthm}
\usepackage{array}\usepackage{algorithm,algorithmic}
\usepackage{booktabs}
\usepackage{graphicx}
\usepackage{epstopdf}\usepackage[utf8]{inputenc}
\usepackage[hyphens,spaces,obeyspaces]{url}
\usepackage[english]{babel}
\usepackage{verbatim} 
\usepackage{lineno,nohyperref}
\usepackage{epstopdf} 
\modulolinenumbers[5]

\newtheorem{definition}{Definition}
\RequirePackage{filecontents}
\usepackage{ragged2e}
\usepackage{url} 
\usepackage[T1]{fontenc}
\usepackage[utf8]{inputenc}
\usepackage{authblk}
\usepackage{microtype} 

\newcommand{\vect}[1]{\boldsymbol{#1}}
\newcommand*{\email}[1]{#1}
\PassOptionsToPackage{hyphens}{url}
\newtheoremstyle{mystyle}
{}
{}                                  
{\itshape}                             
{}                                   
{\bfseries}                         
{.}                                 
{ }                                  
{\thmname{#1}\thmnumber{ #2}\thmnote{ (#3)}}
\newcommand*{\defeq}{\mathrel{\vcenter{\baselineskip0.5ex \lineskiplimit0pt
			\hbox{\scriptsize.}\hbox{\scriptsize.}}}%
	=}
\theoremstyle{mystyle}
\newtheorem{remark}{Remark}
\newtheorem{asu}{Assumption}
\newcounter{subassumption}[asu]
\renewcommand{\thesubassumption}{(\textit{\roman{subassumption}})}
\makeatletter
\renewcommand{\p@subassumption}{\theasu}
\makeatother
\usepackage{amsthm}
\usepackage{xpatch,amsthm}
\makeatletter
\xpatchcmd{\@thm}{\fontseries\mddefault\upshape}{}{}{} 
\makeatother
\newcommand{\subasu}{
	\refstepcounter{subassumption}
	\thesubassumption~\ignorespaces}
\makeatother
\def\BibTeX{{\rm B\kern-.05em{\sc i\kern-.025em b}\kern-.08em
		T\kern-.1667em\lower.7ex\hbox{E}\kern-.125emX}}
\begin{document}
	\title{	Joint Communication, Computation, Caching, and Control in Big Data Multi-access Edge Computing}
\author{Anselme~Ndikumana,~
	Nguyen~H.~Tran,~\IEEEmembership{Member,~IEEE,~}
	Tai~Manh~Ho,~
	Zhu~ Han,~\IEEEmembership{Fellow,~IEEE,~}
	Walid~Saad,~\IEEEmembership{Senior~Member,~IEEE}
	Dusit~Niyato,~\IEEEmembership{Fellow,~IEEE,~}
	and~Choong~Seon~Hong,~\IEEEmembership{Senior~Member,~IEEE}
	\thanks{Anselme Ndikumana, Nguyen H. Tran, Tai Manh Ho, and Choong Seon Hong  are with the Department of Computer Science and Engineering, Kyung Hee University,  Yongin-si, Gyeonggi-do 17104, Rep. of Korea,\\  E-mail:\email{\{anselme, nguyenth, hmtai,  cshong\}@khu.ac.kr}}\\
	\thanks{Zhu Han is with the Electrical and Computer Engineering Department,
		University of Houston, Houston, TX 77004, USA, and the Department of
		Computer Science and Engineering, Kyung Hee University, Yongin-si,
		Gyeonggi-do 17104,  Rep. of Korea, E-mail:\email{	\{zhan2\}@uh.edu}}
	\thanks{Walid Saad is with the Bradley Department of Electrical and Computer Engineering, Virginia Tech, Blacksburg, USA, and the
		Department of Computer Science and Engineering, Kyung Hee University, Rep. of Korea,  E-mail: \email{\{walids\}@vt.edu}}
	\thanks{Dusit Niyato is with the School of Computer Engineering, Nanyang Technological University (NTU), Singapore, E-mail: \email{\{dniyato\}@ntu.edu.sg}}}

\maketitle

\begin{abstract}
	\thispagestyle{empty}
		The concept of multi-access edge computing (MEC) has been recently introduced to supplement cloud computing by deploying MEC servers to the network edge so as to reduce the network delay and alleviate the load on cloud data centers. However, compared to a resourceful cloud, an MEC server has limited resources. When each MEC server operates independently, it cannot handle all of the computational and big data demands stemming from the users’ devices. Consequently, the MEC server cannot  provide significant gains in overhead reduction due to data exchange between users’ devices and remote cloud. Therefore, joint computing, caching,  communication, and control (4C) at the edge with MEC server collaboration is strongly needed for big data applications. In order to address these challenges, in this paper, the problem of joint 4C in big data MEC is formulated as an optimization problem whose goal is to maximize the bandwidth saving while minimizing delay, subject to the local computation capability of user devices, computation deadline, and MEC resource constraints. However, the formulated problem is shown to be non-convex. To make this problem convex, a proximal upper bound  problem of the original formulated problem that guarantees descent to the original problem is proposed. To solve the proximal upper bound problem, a block successive upper bound minimization (BSUM) method is applied. Simulation results show that the proposed approach increases bandwidth-saving and minimizes delay while satisfying the computation deadlines.

\end{abstract}

\IEEEpeerreviewmaketitle

\section{Introduction}
\label{sec:introduction}
\subsection{Background and Motivations} 
In recent years, wireless users have become producers and consumers of contents as their devices are now embedded with various sensors \cite{jin2015quality}, which help in creating and collecting various types of data from different domains such as energy, agriculture, healthcare, transport, security, and smart homes, among others. Indeed, by the year $2020$, it is anticipated that $50$ billion things will be connected to the Internet, which is equivalent to $6$ devices per person on the planet \cite{Cisco}. Therefore, the devices of wireless users will be anywhere, anytime, and connected to anything \cite{zeydan2016big}. This large-scale interconnection of people and things, there will be a tremendous growth of data traffic (from user devices) with different characteristics (unstructured, quasi-structured, and semi-structured) whose scale, distribution, diversity, and velocity fall into a big data framework that requires big data infrastructure and analytics. Since the resources (e.g., battery power, CPU cycles, memory, and I/O data rate) of edge user devices are limited, edge user devices must offload computational tasks and big data to the cloud. However, for effective big data analytics of  delay sensitive and context-aware applications, there is a strong need for low-latency and reliable computation. As such, reliance on a cloud can hinder the performance of big data analytics, due to the associated overhead and end-to-end delays \cite{zeydan2016big, ferdowsi2017deep}.

To reduce end-to-end delay and the need for extensive user-cloud communication, \emph{multi-access edge computing (MEC)} has been introduced by the European Telecommunications Standards Institute (ETSI)  as a supplement to cloud computing and mobile edge computing \cite{hu2015mobile}. MEC  extends cloud computing capabilities by providing IT-based services and cloud computing capabilities at the networks edges. In other words, MEC pushes computation, caching, communication, and control (4C) to the edge of the network \cite{patel2014mobile}. Typically, MEC servers are deployed at the base stations (BSs) of a wireless network (e.g., a cellular network) for executing delay sensitive and context-aware applications in close proximity to the users \cite{semiari2015context, tran2017collaborative, ndikumana2017collaborative}. Therefore, data and computational task offloading  to a nearby MEC server can significantly reduce the end-to-end delay, data exchange between users and the remote cloud, and solve the problem of moving data to the remote cloud and returning computation outputs to the users. In other words, data will be offloaded, processed, analyzed, and cached at the edge of the network , e.g., MEC servers, near where data is created. 

Since offloading requires communication resources, joint optimization of 4C is needed for having an appropriate model that reduces communication and computational delay, while saving backhaul bandwidth. As an example, CCTV security systems use many cameras covering an area for locating people and objects, such as criminals, intruders, missing children, stolen cars, and accidents, where the CCTVs can capture real-time useful videos. Sending real-time streaming videos to a remote cloud for processing and returning the results can strain the network and increase backhaul bandwidth expenses. However, sending real-time streaming videos to nearby MEC servers that can process the videos (e.g., perform object recognition or parsing functions) and return the results can potentially reduce resource usage (e.g., bandwidth) and minimize delay. In addition, the MEC server can cache the videos in its local storage for later use. Therefore, in order to satisfy users' demands, MEC servers located in the same area can \emph{collaborate} through sharing resources.

\subsection{MEC Challenges for Dealing with Big Data} 

The most important challenges that MEC is still facing when dealing with big data and edge analytics are:

\begin{itemize}
	
	\item
	Users offload tasks and corresponding data with varying rates. In other words, data from multiple users may reach MEC servers too rapidly with finite or infinite flow (e.g., streaming data), and this data needs to be processed immediately (e.g., live stream  computation and caching, real-time analytics) \cite{dutta2015big}.  A MEC server will find it challenging to deal with such data due to its scale, diversity, and timeliness. Therefore, for fast, parallel, and distributed processing, MEC servers must support  big data platform and analytics applications for splitting data volume, distributing computations to multiple computing nodes, replicating data partitions, and recovering data when needed. 
	
	\item
	MEC server resources are limited compared to a remote cloud \cite{ahmed2017mobile}.
	Therefore, the MEC server cannot significantly relieve the data exchange between users’ devices and a remote cloud and handle big data MEC efficiently when each MEC server operates independently. Therefore, to reduce the delay, cooperation among MEC servers for resource sharing and optimization of the resource utilization are needed.
	
	\item 
	The integration of MEC with a mobile network environments raise a number of challenges related to the coordination and control of joint communication, computation and caching, and thus, computation and caching depend on the available communication resources. Therefore, joint 4C for big data MEC is needed.  
\end{itemize}

\subsection{Contributions} 
In this work, we address these challenges of joint 4C for big data processing in MEC. The main contributions of this paper are summarized as follows:

\begin{itemize}
	
	\item
	We propose a framework for joint 4C for big data MEC, where big data computation and caching functions are performed at an MEC server instead of being sent to a remote cloud. This allows the reduction of the end-to-end delay and data exchange between users and a remote cloud.
	\item  	
	For satisfying user demands and efficiently executing computational tasks and data caching in big data MEC, we introduce a MEC-based collaboration space or cluster, where MEC servers located in the same cluster collaborate with each other. The aim of the collaboration in MEC is to reduce the backhaul network traffic, minimize delay in coupled 4C, and maximize resource utilization. 
	\item 
	In order to minimize the communication delay among MEC servers and to allow collaboration, inspired by the unsupervised machine learning algorithm called the overlapping k-mean method (OKM)  \cite{cleuziou2008extended}, we propose OKM for collaboration space (OKM-CS). OKM-CS allows each MEC server to participate in more than one collaboration space.  A collaboration space enables collaboration among MEC servers, which is not only based on distance measurements, but also based on available resources.   
	\item
	Within each collaboration space, we formulate a collaborative optimization problem that  maximizes bandwidth savings while minimizing delay, subject to the local computation capabilities of users, computation deadlines, and MEC resources constraints. The formulated problem is shown to be non-convex, and hence, in order to solve it, we propose a proximal upper-bound problem of the original problem and apply the block successive upper bound minimization (BSUM) method, where BSUM is considered as a new and powerful framework for big-data optimization \cite{han2017signal}.
	\item
	Simulation results show that the proposed approach increases bandwidth-saving and minimizes both computation and offloading delay while satisfying user computation deadlines.
\end{itemize}

The rest of the paper is organized as follows. In Section \ref{sec:RW}, we  discuss some related works, while Section \ref{sec:SystemModel} presents the system model. Section \ref{sec:CollaborativeCacheComputation}  discusses in detail  our joint 4C for big data MEC, while Section \ref{sec:PE} provides a performance evaluation. We conclude the paper in Section \ref{sec:Conclusion}.

\section{Literature Review}
\label{sec:RW}
The existing, related works can be grouped into four categories: (i)  big data and caching, (ii) joint caching and computation, (iii) joint caching and communication, and (iv) joint caching, computation, and communication.

\emph{Big data and caching:} In \cite{bacstuug2015big}, the authors proposed a big data framework for mobile network optimization using data from both network features and user features. On the other hand, implementing the big data framework at the network edge can be challenging due to the fact that caching spaces at edge nodes are usually small, which can potentially result in a low hit ratio. To overcome this challenge, in \cite{tran2017collaborative}, the authors highlighted the need of having cooperative caching that allows low latency content delivery. In addition to caching, in \cite{zeydan2016big}, the authors tried to establish connections between big data and caching in 5G wireless networks, where statistical machine learning is applied for estimating content popularity. Other machine learning approaches are surveyed in \cite{chen2017machine}. 

\emph{Joint caching and computation (2C):} In \cite{fan2016terminalbooster}, the authors combined caching and computation at BSs for decreasing delays occurring during communication between applications running on user devices and a remote cloud. They developed a resource management algorithm that guides the BS to jointly schedule computation offloading and data caching allocation. In \cite{lee2017online}, the idea of low-latency computations is explored using the online secretary framework, where the computational tasks are distributed between the edge networks and cloud.
Furthermore, for efficient resource usage at the BS level, in \cite{tran2016collaborative}, the authors proposed a collaborative video caching and processing scheme in which MEC servers can assist each other. They formulated the collaborative joint caching and processing problem as an optimization problem that aims to minimize the backhaul network cost, subject to cache capacity and processing capacity constraints. In \cite{chen2016mobility}, the authors proposed a joint mobility aware caching and small cell base station placement framework. Also, the authors discussed the differences and relationships between caching and computation offloading.

\emph{Joint caching and communication (2C):}  In \cite{jiang2017optimal}, in order to significantly reduce redundant data transmissions and improve content delivery, the authors highlighted the need of having efficient content caching and distribution techniques. They proposed an optimal cooperative content caching and delivery policy in which both femtocell BSs and user equipment participate in content caching. In \cite{hsu2016resource}, the authors studied the problem of resource allocation along with data caching in radio access networks (RANs). They proposed a collaborative framework that leverages device-to-device (D2D) communication for implementing content caching. In \cite{tan2017joint}, a communication framework related to cache-enabled heterogeneous cellular networks with D2D communication was studied. In order to satisfy quality-of-service (QoS) requirements for the users, the authors formulated a joint optimization problem that aims at maximizing the system capacity in which bandwidth resource allocation was considered. The problem of joint caching and communication for drone-enabled systems is also studied in \cite{chen2017caching}.

\emph{ Joint caching, computation, and communication (3C):}
In \cite{zhou2017resource}, the authors combined 3C for designing a novel information centric heterogeneous network framework that enables content caching and computing in MEC. They considered visualized resources, where communication, computing and caching resources can be shared among all users associated with different virtual service providers. Since MEC can enhance the computational capabilities of edge nodes, in \cite{wang2017computation}, the authors formulated a computation offloading decision, resource allocation and data caching framework as an optimization problem in which the total revenue of the network is considered. Furthermore, in \cite{huo2016software}, the authors proposed an energy-efficient framework that considers joint networking, caching, and computing whose goal is to meet the requirements of the next generation of green wireless networks. Moreover, for MEC applications, in \cite{chakareski2017vr}, the authors explored the fundamental tradeoffs between caching, computing, and communication for VR/AR applications. Finally, the work in \cite{cui2017energy} proposed a joint caching and offloading mechanism that considers task uploading and executing, computation output downloading, multi-user diversity, and multi-casting.

In \cite{bacstuug2015big, zeydan2016big, fan2016terminalbooster, zhou2017resource, wang2017computation, huo2016software}, and \cite{cui2017energy}, the authors consider edge caching. However, edge nodes are resources limited as compared to the cloud. Therefore, without cooperation among edge nodes, edge caching can result in a low cache hit ratio. In order to overcome this issue, in \cite{tran2017collaborative} and \cite{tran2016collaborative}, the authors proposed the idea of a collaboration space for edge nodes. However, the works in \cite{tran2017collaborative} and \cite{tran2016collaborative} do not provide any rigorous framework for analyzing the formation of collaboration spaces. Furthermore, a user may request a content format (e.g., avi), which is not available in the cache storage. Instead, the cache storage may have other content formats (e.g., mpeg) of the same content which can be converted to the desired format, by using certain computations, and then transmitted to the requesting user. This process of serving cached content after computation was not considered in \cite{chen2016mobility, jiang2017optimal, hsu2016resource, tan2017joint, chen2017caching}. Finally, the works in \cite{ fan2016terminalbooster,zhou2017resource,wang2017computation, huo2016software} do not take into account any user deadlines for performing computations, which can be impractical.

To this end, our proposed approach will have several key differences from these prior approaches including: \emph{(i)} while many related works (e.g.,\cite{zhou2017resource, wang2017computation, huo2016software, chakareski2017vr, cui2017energy}) focus on 2C and 3C, in our proposed approach,  we combine 4C in big data MEC in which the computation capabilities of the user devices, computation deadline, size of input data, and MEC resource constraints are considered, 
\emph{(ii)} The proposed collaboration between MEC servers, where MEC servers are grouped in collaboration spaces via the OKM-CS algorithm, is new in MEC, and thus is not only based on distance measurements, but also based on the availability of resources,  
\emph{(iii)} Within each collaboration space, for solving  the formulated collaborative optimization problem, we apply the BSUM method, which is not yet utilized in existing MEC solutions. The BSUM method is a novel and powerful framework for big-data optimization \cite{han2017signal}. The BSUM method allows the decomposition of the formulated optimization problem into small subproblems which can be addressed separately and computed in a parallel.

\begin{figure}[t]
	\centering
	\includegraphics[width=0.90\columnwidth]{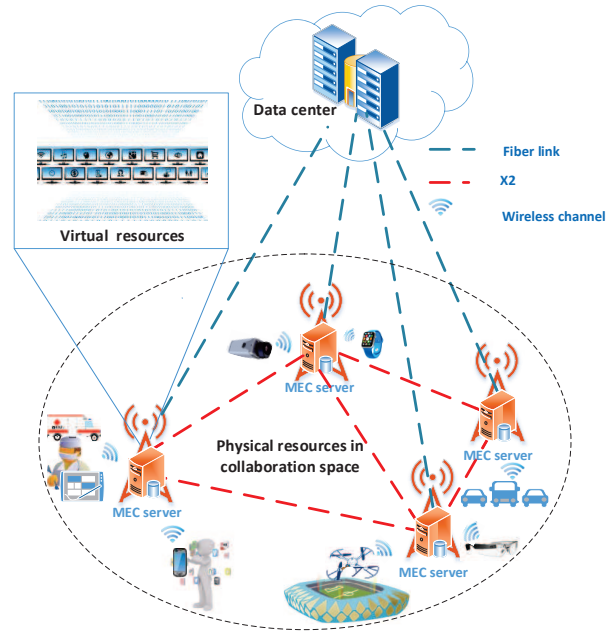}
	\caption{Illustration of our system model.}
	\label{fig:SystemModel}
\end{figure}

\begin{table}[t]
	\caption{Summary of notations.}
	\label{tab:table1}
	\begin{tabular}{ll}
		\toprule
		Notation & Definition\\
		\midrule
		$\mathcal{M}$ & Set of MEC servers, $|\mathcal{M}|= M$\\
		$\mathcal{K}$ & Set of users, $|\mathcal{K}|= K$ \\ 
		$C_m$ & Total cache capacity at MEC server $m \in \mathcal{M}$ \\ 
		$P_m$ & Total computation capacity at MEC server $m \in \mathcal{M}$ \\ 
		$s(d_k)$ & Size of data $d_k$, $\forall k \in \mathcal{K}$\\
		$\tilde{\tau}_{k}$ & Computation deadline, for $k \in \mathcal{K}$ \\
		$\tilde{z}_{k}$ & Computation workload, $\forall k \in \mathcal{K}$ \\
		$\lambda^{d_k}_m$ & Request arrival rate for data $d_k$\\
		&at MEC $m \in \mathcal{M}$ \\
		$l_k$ & Execution latency, $\forall k \in \mathcal{K}$\\
		$E_{k}$ & Computation energy, $\forall  k \in \mathcal{K}$\\
		$\tilde{E_{k}}$ & Available energy in user device $k\in \mathcal{K}$\\
		$ x_k^m$ & Computation offloading decision variable,\\
		&  for $k \in \mathcal{K}$, and $m \in \mathcal{M}$ \\
		$y^{m\rightarrow n}_k  $ & Computation offloading decision variable,\\
		&  for $m,n \in \mathcal{M}$ \\
		$w_m^k$ & Data caching decision variable,  $\forall k \in \mathcal{K}_m, \;m \in \mathcal{M}$ \\
		$\gamma^m_k$& Spectrum efficiency, $\forall k \in \mathcal{K}_m$, and $m \in \mathcal{M}$\\
		$R^m_k$& Instantaneous data rate, $\forall k \in \mathcal{K}_m$, and $m \in \mathcal{M}$  \\
		$\tau_k^{k\rightarrow  m}$& Offloading delay, $\forall k \in \mathcal{K}_m$, and $m \in \mathcal{M}$  \\
		$T_{k}$& Task from user $k \in \mathcal{K}$\\
		$\tau^e_{km}$&Total executing time of offloaded task,\\ & $\forall k \in \mathcal{K}_m$, $m \in \mathcal{M}$\\	
		$\Theta(\vect{x},\vect{y})$& Total delay\\	
		$ \Psi({\vect{x},\vect{y}, \vect{w}})$& Alleviated backhaul bandwidth\\
		\bottomrule
	\end{tabular}
\end{table}
\section{System Model}
\label{sec:SystemModel}

As shown in Fig. \ref{fig:SystemModel}, we consider an MEC network composed of a set $\mathcal{M}$ of MEC servers, each of which  is attached to one BS. Unless stated otherwise, we use the terms \enquote{MEC server} and \enquote{BS} interchangeably. 


Each MEC server collaborates with other MEC servers by sharing resources. Therefore, we group the BSs into \emph{collaboration spaces (i.e., clusters)}. Unless stated otherwise, we use the terms \enquote{collaboration space} and \enquote{cluster} interchangeably. In order to minimize the communication delay among MEC servers, our clustering process for BSs is based on proximity (distance) measurements, where BSs that are close enough will be grouped in the same cluster. Moreover, in our collaboration space, we focus on geographic space coverage rather than geographical space partitioning. As an example, some MEC servers in the hotspot area may want to collaborate with MEC servers not in the hotspot. To achieve this objective, we consider an overlapping clustering method that allows one BS to belong to more than one cluster and to share resources not only based on distance measurements, but also based on resource availability and utilization.

For creating collaboration spaces, we propose OKM for collaboration space (OKM-CS), which is a modified version of the standard OKM algorithm \cite{cleuziou2008extended}. The merit of the OKM algorithm lies in its elegant simplicity of implementation over other overlapping methods such as weighted OKM (WOKM), overlapping partitioning cluster (OPC), and multi-cluster overlapping k-means extension (MCOKE) \cite{baadel2016overlapping}. OKM-CS is described in Section \ref{sec:CollaborativeCacheComputation}, Algorithm \ref{algo:OKM}. 

In a collaboration space, each MEC server $m$ has both caching and computational resources that are divisible. We let $C_m$ and $P_m$ be, respectively, the cache capacity and computational capacity of MEC server $m$. In any given collaboration space, MEC servers can exchange data and tasks based on their available resources. Moreover, we assume that the MEC servers within a collaboration space belong to the same mobile network operator (MNO), and this MNO has a total storage capacity $C$, and a total computation capacity $P$.  The total cache storage pool for the MNO in a collaboration space is given by:
\begin{equation}
\begin{aligned}
C= \sum_{m \in \mathcal{M}} C_m,
\end{aligned}
\label{eq:totalcache}
\end{equation} 
while the computation pool of the MNO is given by:
\begin{equation}
\begin{aligned}
P= \sum_{m \in \mathcal{M}} P_m.
\end{aligned}
\label{eq:computation}
\end{equation}

We assume that each MEC server $m $ uses a resource allocation table (RAT) for keeping track of the available resources in the collaboration space, including CPU utilization, RAM, and storage capacity. In order to facilitate joint 4C in big data MEC, in collaboration space, MEC servers exchange RAT updates. However, for the resources that are not available in a collaboration space, MEC server $m$ forwards the associated requests to the remote data center (DC). Therefore, for effective resource utilization, resources are sliced for being allocated to multiple users.

We consider a set $\mathcal{K}$ of users, where each user $k \in \mathcal{K}$ is connected to its nearest BS, referred to as its home BS. The set of users connected to the same BS $m\in M$ is denoted by a subset $\mathcal{K}_m \subset \mathcal{K}$. We assume that the user devices have limited resources for both computation and caching. Therefore, instead of sending resource demands to the DC, based on user demands, MEC servers can provide computation and storage resources to the users. As an example, drones in professional sports activities can cover the event scenes and send live stream videos to their nearest MEC server $m$ for live stream caching, processing, and distribution. Based on the network conditions, user demands, and device capabilities, the cached data can be served as is or after computation (e.g., video transcoding).

In our model, each user device $k \in \mathcal{K}$ has an application that needs to use computation and caching resources, such as augmented reality, online gaming, crowdsensing, image processing, or CCTV video processing.

We consider a binary task offloading model in which a task is a single entity that is either computed locally at a user device or offloaded to the MEC server. For each user $k$, we define a task $T_{k} = (s(d_k),\tilde{\tau}_{k}, \tilde{z}_{k}),\ \forall k \in \mathcal{K}$, where $s(d_k)$ is the size of data $d_k$ from user $k$ in terms of the bits that are needed as an input of computation, $\tilde{\tau}_{k}$ is the task computation deadline, and $\tilde{z}_{k}$ is the computation workload or intensity in terms of CPU cycles per bit. Furthermore, we assume that the resource demands of different users are independent. 

\begin{figure}[t]
	\centering
	\includegraphics[width=0.7\columnwidth]{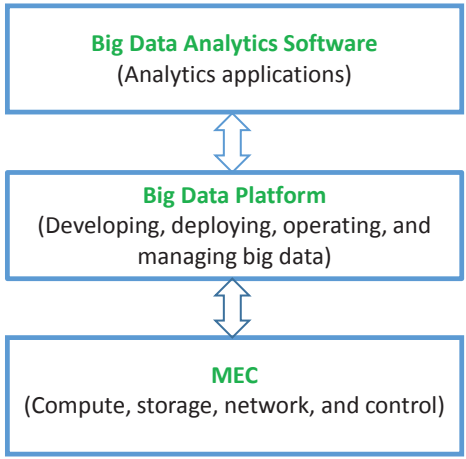}
	\caption{Illustration of big data MEC.}
	\label{fig:big-data}
\end{figure}

In order to satisfy user demands, as depicted in Fig \ref{fig:big-data}, we consider each MEC server to be a small big data infrastructure that supports the big data cloud requirements defined in \cite{kune2016anatomy}, including ($i$) Easy setup of virtual machines, mounting file systems, and deployment of big data platform and analytics software such as Hadoop, Spark, Storm, and Splunk; ($ii$)	Dynamic management of computation, storage, and network resources, either on physical or virtual environments; ($iii$) Elasticity and scalability in computation, storage and network resources allocation; ($iv$) Development, deployment  and utilization of big data analytics with fast access to data and computing resources; and ($v$) Support for multi-dimension data handling, where data may reach the MEC server in different forms and characteristics.

\section{Proposed Joint Communication, Computation, Caching, and Control}
\label{sec:CollaborativeCacheComputation}

In this section, we describe, in detail, our proposed approach for joint communication, computation, caching, and  distributed control in big data MEC, where MEC server resources are virtualized and shared by multiple users. Resource demands that are not satisfied at one MEC server can be satisfied by any other MEC server in the same collaboration space.

\subsection{Collaboration Space}
\label{subsec: collaboration-space} 

For forming collaboration spaces, we propose OKM-CS. OKM-CS seeks to cluster the BSs into $r$ clusters such that the below objective function is minimized:
\begin{equation}
\begin{aligned}
\mathcal{I}(\{\mathcal{M}_i\}_{i=1}^r)=\sum_{i=1}^r   \sum_{m \in \mathcal{M}_i}\lVert m-\Phi(m)\rVert^2,
\end{aligned}
\label{eq:OKM_objective}
\end{equation}
where $\mathcal{M}_i\subset \mathcal{M}$ represents the $i^{th}$ cluster. \textcolor{black}{ Furthermore, as defined in \cite{cleuziou2008extended}, $\Phi(m)$ defines the average of centroids ($m_{c_i}$) of the clusters to which the BS $m$ belongs, and is given by:
	\begin{equation}
	\Phi(m) =\frac{\sum_{m_{c_i} \in \mathcal{A}_i^m} m_{c_i}}{|\mathcal{A}_i^m|},
	\end{equation}
	where $\mathcal{A}_i^m$ defines multi-assignment for BS $m$: $\{m_{c_i}| m \in \mathcal{M}_i\}$, which means that $\mathcal{A}_i^m$ is a set of all centroids $m_{c_i}$ for which  $m \in \mathcal{M}_i$.
	In other words,  each BS $m$ belongs to at least one cluster, where $\bigcup_{i=1}^r\mathcal{M}_i=\mathcal{M}$ represents the total coverage.}

The original OKM algorithm randomly chooses  $r$ clusters. However, in OKM-CS for 4C, the number of clusters is chosen  based on the network topology, which is known a priori by the  MNO. OKM-CS  is presented in Algorithm \ref{algo:OKM}. 

Algorithm \ref{algo:OKM} starts with an initial set of $r$ clusters and  centroid $\{m_{c_i}^{(0)}\}_{i=1}^r$, and derives new coverage $\{\mathcal{M}_i^{(0)}\}_{i=1}^r$. Then, it iterates by computing new assignments and new centroids $\{m_{c_i}^{(t+1)}\}_{i=1}^r$ leading to the new coverage $\{\mathcal{M}_i^{(t+1)}\}_{i=1}^r$. The iterative process continues until the convergence criterion on $\mathcal{I}(\{\mathcal{M}^{(t)}_i\}_{i=1}^r)-\mathcal{I}(\{\mathcal{M}^{(t+1)}_i\}_{i=1}^r)<\epsilon$) is satisfied, where  $\epsilon$ is a small positive number. Furthermore, since our focus is on the collaboration among the MEC servers in the same collaboration space, for brevity, hereinafter, we omit the subscript on $\mathcal{M}_i$ and analyze 4C for one collaboration space.

\begin{algorithm}[t]
	\caption{: OKM for collaboration space (OKM-CS)}
	\label{algo:OKM}
	\begin{algorithmic}[1]
		\STATE{\textbf{Input:} $\mathcal{M}$: A set of BSs with their coordinates, \\ $t_m$: Maximum number of iterations, $\epsilon>0$; }
		\STATE{\textbf{Output:} $\{\mathcal{M}_i^{(t+1)}\}_{i=1}^r$ : Final cluster coverage of BSs;}
		\STATE {Choose $r$  and initial clusters with $\{m_{c_i}^{(0)}\}_{i=1}^r$} centroid;
		\STATE{ For each BS $m$, compute the assignment\\ $\mathcal{A}^{m(0)}_i$ by assigning bs $m$ to centroid $\{m_{c_i}^{(0)}\}_{i=1}^r,$ and derive initial coverage $\{\mathcal{M}_i^{(0)}\}_{i=1}^r$, such that $\mathcal{M}_i^{(0)}=\{m|m_{c_i}^{(0)} \in \mathcal{A}^{m(0)}_i \}$;}
		\STATE{Initialize $t=0$;}
		\STATE{For each cluster $\mathcal{M}_i^{(t)}$, compute the
			new centroid,\\ $ m_{c_i}^{(t+1)}$ by grouping $\mathcal{M}_i^{(t)}$; }
		\STATE{ For each BS $m$ and assignment $\mathcal{A}^{m(t)}_i$, compute new assignment  $\mathcal{A}^{m(t+1)}_i$ by assigning bs $m$ to centroid $\{m_{c_i}^{(t+1)}\}_{i=1}^r$ and derive new coverage $\{\mathcal{M}_i^{(t+1)}\}_{i=1}^r$;}
		\STATE{If equation (\ref{eq:OKM_objective}) does not converge or $t_m>t$ or $ \mathcal{I}(\{\mathcal{M}^{(t)}_i\}_{i=1}^r)-\mathcal{I}(\{\mathcal{M}^{(t+1)}_i\}_{i=1}^r)>\epsilon$, set $t=t+1$, restart from Step $6$. Otherwise, stop and consider $\{\mathcal{M}_i^{(t+1)}\}_{i=1}^r$ as the final clusters.}
	\end{algorithmic}
\end{algorithm}

In a collaboration space, for the MEC resources, each user $k$ must submit a task demand $T_{k}$ to its MEC server $m$.  Then, the MNO maps the demands into the resource allocation that each user $k$ requires. Therefore, to help the users prepare their demands, the MNO advertises the resources available to them as well as the sum of the demands placed in the collaboration space. However, the MNO does not reveal the demands of the users to each other.

We use $v_{km}(c_{dk},p_{km}, R^m_k)$ to represent the resource allocation function for each user $k$ at MEC server $m$, where $c_{dk}$ is used to denote the caching resource allocation for user data of size $s(d_k)$ (i.e., $c_{dk}=s(d_k)$),  $ p_{km}$ is used to denote the computational resource allocation, and $R^m_k$ is used to denote the communication resource allocation.

The MNO allocates resources based on weighted proportional allocation \cite{nguyen2011weighted}, which is practical in systems such as 4G and 5G cellular networks \cite{mosleh2016proportional, lei2015joint}. Each user $k$ receives a fraction of the resources at the MEC server $m$ based on its demand. Furthermore, when $\tilde{\tau}_{k}=0$ and $\tilde{z}_{k}=0$, we consider that the user needs only communication resources for offloading data $d_k$ and caching resources for caching its data. Therefore, an MEC server caches data $d_k$, and waits for the data to be requested later, where $d_k$  can be served as is or after computation. 
However, when $s(d_k) \neq 0$, $\tilde{\tau}_{k}\neq 0$, and $\tilde{z}_{k} \neq 0$, the MEC server computes, caches the output data of $d_k$, and returns the computation output to user $k$.

\subsection{Communication Model}
\label{subsec:communication_model} 

\begin{figure}[t]
	\centering
	\includegraphics[width=1.0\columnwidth]{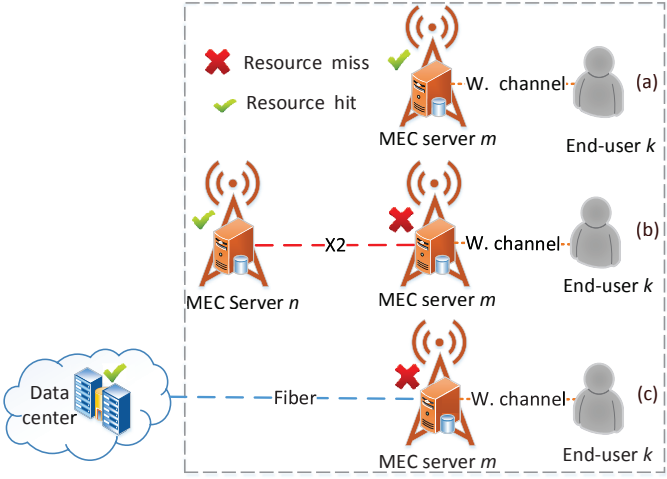}
	\caption{Collaboration space for MEC with three typical scenarios (a), (b), and (c), which are explained in Section \ref{subsec:communication_model}.}
	\label{fig:content_distribution}
\end{figure}

To offload a task from a user to the MEC server, the network will incur a communication cost, where the user can offload tasks through use of the communication scenarios, shown in Fig. \ref{fig:content_distribution} and explained next.

\emph{Scenario (a):} For the resources available at BS $m \in \mathcal{M}$, user  $k \in \mathcal{K}$ obtains resources from its MEC server over a wireless channel. We define $ x_k^m\in \{0,1\}$ as a computation offloading decision variable, which indicates whether or not user $k$ offloads the task to its home MEC server $m$ via a wireless channel (denoted by W. channel in Fig. \ref{fig:content_distribution}).
\begin{equation}
\setlength{\jot}{10pt}
x_k^m=
\begin{cases}
1,\; \text{if $T_k$ is offloaded from user $k$ to BS $m$},\\
0, \;\text{otherwise.}
\end{cases}
\end{equation}

Therefore, the spectrum efficiency for user device $k$ is expressed as:
\begin{equation}
\label{eq:SINR}
\begin{aligned}
\gamma^m_k = \log_2\left(1 + \frac{\rho_k |G^m_k|^2}{\sigma_k^2}\right),  \;\forall k \in \mathcal{K},\; m \in \mathcal{M},
\end{aligned}
\end{equation}
where $\rho_k$ is the transmission power of user device $k$, $|G^m_k|^2$ is the channel gain between user device $k$ and BS $m$, and $\sigma_k^2$ is the power of the Gaussian noise at user $k$.

The instantaneous data rate for user device $k$ is given by:
\begin{equation}
R^m_k= x_k^ma_k^m B_m\gamma^m_k, \forall k \in \mathcal{K},\; m \in \mathcal{M}, 
\label{eq:instantaneous_data}
\end{equation} 
where each user $k$ of BS $m$ is allocated a fraction $a_k^m$ ($ 0\leq a_k^m \leq 1$) of bandwidth $B_m$. We assume that the spectrum of the MNO is orthogonal so that there is no interference among the users. Furthermore, we assume that user demand for offloading will only be accepted if there is enough spectrum resources to satisfy its demand.

Based on the instantaneous data rate, as defined in \cite{mao2017survey}, the transmission delay for offloading a task from user $k$ to the MEC server $m$ is expressed as:

\begin{equation}
\tau_k^{k \rightarrow m}=\frac{x_k^m s(d_k)}{R^m_k}, \; \forall k \in \mathcal{K}_m,
\end{equation}
where $\mathcal{K}_m$ is a set of users served by BS $m$.

\emph{Scenario (b):} 
When the MEC server $m$ has insufficient resources to satisfy the user demand, after checking its RAT, BS $m$ forwards a request to another BS $n$ via an X2 link \cite{x2}, in the collaboration space, which has enough resources. Therefore, users can get the resources from different MEC servers with different delay costs.

We define $y^{m\rightarrow n}_k$ as a decision variable, which indicates whether or not the task of user $k$ is offloaded from BS $m$, as follows: 
\begin{equation}
\setlength{\jot}{10pt}
y^{m\rightarrow n}_k =
\begin{cases}
1,\; \text{if $T_k$ of user $k$ is offloaded from BS $m$}
\\ \; \; \; \;\text{to a neighbor BS $n$,}\\
0,\;\text{otherwise.}
\end{cases}
\end{equation}

We denote by $\tau^{m\rightarrow n}_k$ the offloading delay between BS $m$ and BS $n$, which is given as follows:
\begin{equation}
\tau^{m\rightarrow n}_k=\frac{\sum_{k \in \mathcal{K}_m}y^{m\rightarrow n}_k s(d_k)}{\Gamma^n_m}, \;\forall m,\;n \in \mathcal{M},
\end{equation}
where $\Gamma^n_m$ is the X2 link capacity between BS $m$ and BS $n$.

\emph{Scenario (c):}
When the resources are not available in the whole collaboration space, BS $m$ forwards the request to the remote cloud  through a wired backhaul link. 

We define $y^{m\rightarrow DC}_k$ as a decision variable that indicates whether or not the task of user $k$ is offloaded  by BS $m$ to  the DC, as follows:
\begin{equation}
\setlength{\jot}{10pt}
y^{m\rightarrow DC}_k=
\begin{cases}
1,\; \text{if $T_k$ is offloaded from BS $m$ to the DC},\\
0, \;\text{otherwise.}
\end{cases}
\end{equation}

We define $\tau^{m\rightarrow DC}_k$ as the offloading delay between BS $m$ and DC $l$, where $\tau^{m\rightarrow DC}_k$ is given by:
\begin{equation}
\tau^{m\rightarrow DC}_k=\frac{\sum_{k \in \mathcal{K}_m}y^{m\rightarrow DC}_k s(d_k)}{\Omega^{DC}_m}, \;\forall m,\; n \in \mathcal{M},
\end{equation}
where $\Omega^{DC}_m$ is the link capacity between  MEC server $m$ and remote DC. 

\subsection{Computation Model} 
\label{subsec: computation_model}

\subsubsection{ Local Computation at User Device} 
In our model, we assume that each user device $k \in \mathcal{K}$ has a task $T_{k}$ that needs to use the local computation resource $P_k$ of device $k$. Therefore, the computation of task $T_{k}$ requires CPU energy $E_{k}$, where the energy consumption of CPU computation at user $k$, as defined in \cite{mao2017survey}, is expressed as: 

\begin{equation}
E_{k}=  s(d_k)\nu \tilde{z}_{k}P_{k}^2,  \;k \in \mathcal{K},
\end{equation}
where $\nu$ is a constant parameter that is related to the CPU hardware architecture. 

In addition to the CPU energy consumption, the computation of task $T_{k}$ requires execution time $l_k$. Therefore, as defined in \cite{mao2017survey}, the execution latency for task $T_{k}$ at user device $k$ is given by:

\begin{equation}
\label{eq:compution_time}
l_k=\frac{s(d_k)\tilde{z}_{k} }{P_{k}}.
\end{equation}

However, when $l_k> \tilde{\tau}_{k}$, $ \tilde{z}_{k}>P_{k}$, or $ E_{k}>\tilde{E_{k}}$, where $\tilde{E_{k}}$ is the actual available energy at user device $k \in \mathcal{K}$,  device $k$ does not have enough energy or computation resources to meet the computation deadline, and  thus,  user $k$ can keep the computational task until the resources become available for local computation via its device. Therefore, we define $\alpha_{k}\in \{0,1\}$ as a user device status parameter for computing task $T_{k}$, where the value of $\alpha_{k}$ can be set as follows: 
\begin{equation}
\setlength{\jot}{10pt}
\alpha_{k} =
\begin{cases}
0,\; \text{if $\tilde{z}_{k}>P_{k}$, or $l_k>\tilde{\tau}_{k}$, or $ E_{k}>\tilde{E_{k}}$},\\
1, \;\text{otherwise.}
\end{cases}
\end{equation}

From the value of $\alpha_{k}$, the total local execution time $\tau^\textrm{loc}_{k}$ of task $T_{k}$ at user device $k$ is given by:
\begin{equation}
\setlength{\jot}{10pt}
\tau^\textrm{loc}_{k} =
\begin{cases}
l_k\;, \text{if $\alpha_{k}=1$, and  $x_k^m=0$},\\
l_k+ \varphi_k\;, \text{if $\alpha_{k}=0$, and  $x_k^m=0$},\\
0, \; \text{if $\alpha_{k}=0$, and  $x_k^m=1$},\\
\end{cases}
\end{equation}
where $\varphi_k$ is the average waiting time of task $T_{k}$ until it is locally executed by device $k$.

Each user $k \in \mathcal{K}$ can compute its task $ T_{k}$ locally on its device, when the device has enough resources, in terms of the CPU cycles, energy, or memory, whenever the user device status parameter is $\alpha_{k}=1$. However, if user $k$ decides not to offload its task to an MEC server, it will experience a computational delay $\tau^\textrm{loc}_{k}$. Therefore, if user $k$ cannot keep a given computational task for the future and $l_k> \tilde{\tau}_{k}$, $ \tilde{z}_{k}>P_{k}$, $E_{k}>\tilde{E_{k}} $ ($\alpha_{k}=0$), then this user $k$ can offload the task to MEC server $m$.

\subsubsection{Computation at MEC Server} 
We consider $P_m$ as the available computational resources at the MEC server $m \in \mathcal{M}$. Furthermore, we define $y_k^{k\rightarrow m}\in \{0,1\}$ as a decision variable, which indicates whether or not MEC server $m$ has to compute the task $T_k$ offloaded by user $k$, where $y_k^{k\rightarrow m}$ is given by:

\begin{equation}
\setlength{\jot}{10pt}
y_k^{k\rightarrow m} =
\begin{cases}
1,\; \text{if $T_k$ offloaded by user $k$}\\
\; \; \; \;\text{is computed at BS $m$},\\
0, \;\text{otherwise.}
\end{cases}
\end{equation}
The computation allocation $p_{km}$ at BS $m$ can be calculated as follows:

\begin{equation}
p_{km}=P_m\frac{\tilde{z}_{k}}{\sum_{g\in \mathcal{K}_m}\tilde{z}_{g}},\; \forall k  \in \mathcal{K}_m,\; m \in \mathcal{M}.
\end{equation}

At each MEC server $m$, the total computation allocations must satisfy:

\begin{equation}
\sum_{k\in \mathcal{K}_m}x_k^mp_{km}y_k^{k\rightarrow m}\leq P_m,\; \forall  m \in \mathcal{M}.
\end{equation}

The execution latency $l_{km} $ for task  $ T_{k}$ at  MEC server $m$ is given by:
\begin{equation}
\label{eq:compution_time_mec}
l_{km} =\frac{s(d_k) \tilde{z}_{k}}{p_{km}}.
\end{equation}
Therefore, the total execution time for task $T_k$ that was offloaded by user $k$ at MEC server $m$ is given by:
\begin{equation}
\tau^e_{km}= \tau^{k\rightarrow  m}_k + l_{km},\; \forall k \in \mathcal{K}_m,\; m \in \mathcal{M}.
\end{equation}

However, if  $\tilde{z}_{k}>p_{km} $ or $\tau^e_{km} > \tilde{\tau}_{k}$ (i.e., MEC server $m$ does not have enough computational resources to meet the computation deadline),  MEC server $m$ checks its RAT and offloads the task to any MEC server $n$ that has enough resources to satisfy the demand. Here, $l_{kn}$ is the execution latency for task  $ T_{k}$ at  MEC server $n$ and can be calculated based on (\ref{eq:compution_time_mec}). Therefore, the total execution time for a  task offloaded by user $k$ to MEC server $n$ becomes:
\begin{equation}
\tau^e_{kmn}=\tau_k^{k \rightarrow m} + \tau^{m\rightarrow n}_k + l_{kn},\; \forall k \in \mathcal{K}_m, \text{ and } m,n \in \mathcal{M}.
\end{equation}

When there are no available resources in a collaboration space, MEC server $m$ offloads the task to the DC. Therefore, the total execution time for task $T_k$  offloaded by user $k$ at DC becomes:
\begin{equation}
\tau^e_{kmDC}=\tau_k^{k \rightarrow m} + \tau^{m\rightarrow DC}_k + l_{kDC},\; \forall k \in \mathcal{K}_m, \text{ and } m \in \mathcal{M},
\end{equation}
where $l_{kDC}$ can be calculated from (\ref{eq:compution_time_mec}). Furthermore, we find the total offloading and computation latency $\tau^\textrm{off}_{k}$ for task $T_k$  offloaded by user $k$ as follows:
\begin{equation}
\begin{aligned}
\tau^\textrm{off}_{k}=y_k^{k\rightarrow m}\tau^e_{km} + \sum_{n\in \mathcal{M}}y^{m\rightarrow n}_k\tau^e_{kmn}+ y^{m\rightarrow DC}_k\tau^e_{kmDC},\\ \forall k \in \mathcal{K}_m, \text{ and } m \in \mathcal{M}.
\end{aligned} 
\end{equation}

In order to ensure that task $T_k$ is executed at only one location, i.e., computed locally at a user device, at one of the MEC servers, or at the remote cloud, we impose the following constraints, $\forall m \in \mathcal{M}$:
\begin{align}
\setlength{\jot}{10pt}
(1-x^m_k) + x^m_k(y_k^{k\rightarrow m}+\sum_{n\in \mathcal{M}}y^{m\rightarrow n}_k + y^{m\rightarrow DC}_k)= 1,
\label{eq:offloading_constraint}
\end{align}
\begin{align}
\max \{y_k^{k\rightarrow m}, y^{m\rightarrow n}_k, y^{m\rightarrow DC}_k,\forall n  \}\leq x^m_k, \;\forall k\in \mathcal{K}_m.
\label{eq:green-constraint-59}
\end{align}

\subsection{Caching Model}
\label{subsec:caching_model}

For an offloaded task $ T_{k}$, when $\tilde{\tau}_{k}=0$ and $\tilde{z}_{k}=0$, MEC server $m$ caches data $d_k$. Based on user $k$'s demand  $\lambda^{d_k}_m$ for data $d_k$ that reaches each MEC server $m$, $d_k$ can be retrieved from the cache storage. Here, using the idea of a cacheable task that was defined in \cite{elbamby2017proactive}, we assume that all tasks are cacheable. However, due to the limited cache capacity, the MNO needs to evict from cache the least frequently reused data in order to make room for new incoming data that needs to be cached. During data replacement, the MEC server starts replacing the least frequently reused data based on the number of requests  $\lambda^{d_k}_m$ that are satisfied by the MEC severs, i.e., the number of cache hits. Here, the well-known least frequently used (LFU)  cache replacement policy \cite{lee2001lrfu} \cite{ndikumana2016scalable} is applied.

We let $w_m^k\in \{0,1\}$ be the decision variable that indicates whether or not MEC server $m$ has to cache data $d_k$ of user $k$, where $w_m^k$ is given by:
\begin{equation}
\setlength{\jot}{10pt}
w_m^k=
\begin{cases}
1,\; \text{if MEC server $m \in \mathcal{M}$ caches the data $d_k$},\\
0, \;\text{otherwise.}
\end{cases}
\end{equation}

We let $C_m$ be the cache capacity available at any MEC server $m$. Therefore, the total allocation of caching resources at MEC server $m$ must satisfy:
\begin{equation}
\begin{aligned}
\label{eq:cache_allocation}
\left(\sum_{k\in \mathcal{K}_m} y_k^{k \rightarrow m} + \sum_{n\neq m\in \mathcal{M}}\sum_{k\in \mathcal{K}_n}  y^{n\rightarrow m}_k\right) w_m^k s(d_k)\leq C_m, \\ \forall  m \in \mathcal{M}.
\end{aligned}
\end{equation}

When MEC server $m$ does not have enough cache storage to cache data $d_k$,  MEC server $m$ checks its RAT, and offloads $d_k$ to MEC server $n$ in the collaboration space (if MEC server $n$ has enough cache storage to satisfy the demand) or forwards the request to the DC. When data $d_k$  is requested at MEC server $m$, it will either be  served from a cache in the collaboration space or forwarded to the DC if $d_k$ is not cached in the collaboration space.

\subsection{Distributed Optimization Control} 
\label{subsec:control_model}
Next, we propose a distributed control model, which is based on a distributed optimization that coordinates and integrates the communication, computation, and caching models defined in the previous sections. 

In the distributed control model, we maximize the backhaul bandwidth saving (minimize the backhaul bandwidth) by reducing the data exchange between MEC servers and remote DC, i.e., increasing the cache hits. Therefore, we adopt the caching reward defined in \cite{wang2017computation} as the amount of  saved backhaul bandwidth given by:
\begin{equation}
\begin{aligned}
\Psi({\vect{x},\vect{y}, \vect{w}})= \sum_{m\in \mathcal{M}}\sum_{k\in \mathcal{K}_m} s(d_k)\lambda^{d_k}_m  x^m_k(y_k^{k\rightarrow m}  w_m^{k} \\+\sum_{n\in \mathcal{M}}y^{m\rightarrow n}_k w_n^k),
\end{aligned}
\label{eq:caching_reward}
\end{equation} 
where the requests for data $d_k$ arrive at BS $m$ with arrival rate  $\lambda^{d_k}_m$.

Here, we consider the total delay as the  total amount of time that task $T_{k}$ takes to be completely computed (offloading delay included). For the computation cost, if user $k$ computes its task locally, then the computational delay cost of $\tau^\textrm{loc}_k$ is incurred. On the other hand, when user $k$ decides to offload the computational task to an MEC server, a total offloading and computation delay of $\tau^\textrm{off}_{k}$ is incurred. In order to minimize both computation  delay costs ($\tau^\textrm{loc}_{k}$ and $\tau^\textrm{off}_{k}$), we formulate the total delay $\Theta(\vect{x},\vect{y})$ for the tasks computed locally at user devices, or in the MEC collaboration space, or at the remote cloud as follows:
\begin{equation}
\begin{aligned}
&\Theta(\vect{x},\vect{y})=\sum_{m\in \mathcal{M}}\sum_{k\in \mathcal{K}_m} (1-x_k^m)\tau^\textrm{loc}_{k} +x_k^m\tau^\textrm{off}_{k}.
\end{aligned}
\label{eq:total_compution_cost}
\end{equation}

\subsubsection{Problem Formulation} 
We formulate the joint 4C in big data MEC as an optimization problem that aims at maximizing bandwidth saving while minimizing delay, subject to the local computation capabilities of user devices, and MEC resource constraints as follows:
\begin{subequations}\label{eq:problem_formulation}
	\begin{align}
	&\underset{\vect{x}, \vect{y}, \vect{w}}{\text{min}}\ \  \Theta(\vect{x},\vect{y})-\eta \Psi({\vect{x},\vect{y}, \vect{w}})
	\tag{\ref{eq:problem_formulation}}\\
	&\text{subject to:}\nonumber\\
	& \sum_{k\in \mathcal{K}_m}x_k^ma^m_{k}\leq 1, \;  \forall m \in \mathcal{M},\label{first:a}\\
	&\sum_{k\in \mathcal{K}_m}x_k^mp_{km}y_k^{k\rightarrow m}\leq P_m,\forall  m \in \mathcal{M}, \label{first:b}\\
	&x_k^m (\sum_{k\in \mathcal{K}_m} y^{k \rightarrow m} + \sum_{n\neq m\in \mathcal{M}}\sum_{k\in \mathcal{K}_n}  y^{n\rightarrow m}_k) w_m^k s(d_k)\leq C_m, \label{first:c}\\
	&(1-x^m_k) + x^m_k(y_k^{k\rightarrow m}+\sum_{n\in \mathcal{M}}y^{m\rightarrow n}_k + y^{m\rightarrow DC}_k)= 1, \label{first:d}\\
	&\max \{y_k^{k\rightarrow m}, y^{m\rightarrow n}_k, y^{m\rightarrow DC}_k,\forall n  \}\leq x^m_k,\label{first:e}
	\end{align}
\end{subequations}
where $\eta>0$ is the weight parameter, which is typically used in multi-objective optimization \cite{deb2014multi}.

The constraint in (\ref{first:a}) guarantees that the sum of spectrum allocation to all users has to be less than or equal to the total available spectrum at each BS $m$. The constraints in (\ref{first:b}) and (\ref{first:c}) guarantee that the computation and cache resources allocated to users at each MEC server $m$ do not exceed the computation and caching resources. The constraints in (\ref{first:d}) and (\ref{first:e}) ensure that the task $T_k$ has to be executed at only one location, i.e., no duplication. Furthermore, in order to simplify the notation, we define the new objective function:
\begin{equation}
\begin{aligned}
\mathcal{B}({\vect{x},\vect{y},\vect{w}})\defeq \Theta(\vect{x},\vect{y})-\eta\Psi({\vect{x},\vect{y}, \vect{w}}).
\end{aligned}
\label{eq:transformed_equation}
\end{equation}

The above optimization problem in (\ref{eq:transformed_equation}) is difficult to solve due to its non-convex structure. Therefore, to make it convex, we use BSUM method  described in below Section \ref{subsubsec:BSUM}. 

\subsubsection{Overview of BSUM Method} 
\label{subsubsec:BSUM}
BSUM is a distributed algorithm that allows parallel computing. The advantages of BSUM over centralized algorithms reside in both solution speed and problem decomposability \cite{han2017signal}. Therefore, for introducing  BSUM \cite{hong2016unified} in its standard form, we consider the following function as a block-structured optimization problem:
\begin{equation}
\begin{aligned}
\underset{\vect{x}}{\text{min}} \;g({\vect{x}_1,\vect{x}_2,\ldots,\vect{x}_J}), \; s.t. \; \vect{x}_j \in \mathcal{Z}_j,\;\forall j \in \mathcal{J}^t,\; j=1,\ldots,J,
\end{aligned}
\label{eq:optimization_bsum}
\end{equation} 
where $\mathcal{Z}:=\mathcal{Z}_1  \times \mathcal{Z}_2 \times \cdots \mathcal{Z}_J$, $g(.)$ is a continuous function, and $\mathcal{J}^t$ is the set of indexes. For  $j=1,\ldots,J$, we consider $\mathcal{Z}_j$ as a closed convex set, and  $\vect{x}_j$ as a block of variables. By applying BCD, at each iteration $t$, a single block  of variables is optimized by solving the following problem:
\begin{equation}
\begin{aligned}
\vect{x}^t_j \in \underset{\vect{x}_j \in \mathcal{Z}_j}{\text{argmin}}\; g({\vect{x}_j, \;\vect{x}^{t-1}_{-j}}),
\end{aligned}
\label{eq:optimization_BCD}
\end{equation}
where $\vect{x}^{t-1}_{-j}:=(x^{t-1}_{1},\ldots, x^{t-1}_{j-1},x^{t-1}_{j+1},\ldots, x^{t-1}_{j}$), $\vect{x}^{t}_{k}=\vect{x}^{t-1}_{k}$ for $j\neq k$.

Both problems in (\ref{eq:optimization_bsum}) and (\ref{eq:optimization_BCD}) are difficult to solve, especially when (\ref{eq:optimization_bsum}) is a non-convex function, and block coordinate descent (BCD) does not always guarantee convergence. Therefore, with BSUM, at a given feasible point $\vect{y} \in \mathcal{Z}$, we can introduce the proximal upper-bound function $h(\vect{x}_j,\vect{y})$ of $g(\vect{x}_j,\vect{y}_{-j})$. The most commonly used schemes for choosing the  proximal upper-bound function are proximal upper-bound, quadratic upper-bound, linear upper-bound and Jensen's upper-bound \cite{hong2016unified}. The proximal upper-bound function $h(\vect{x}_j,\vect{y})$ must satisfy following Assumption $1$:
\begin{asu}\label{1} \textrm{We make the following assumptions:}\\
	\subasu \label{1A} $h(\vect{x}_j,\vect{y})=g(\vect{y})$,\\
	\subasu \label{1B} $h(\vect{x}_j,\vect{y})>g(\vect{x}_j,\vect{y}_{-j})$,\\
	\subasu \label{1C} $h'(\vect{x}_j,\vect{y};\vect{q}_j)|_{\vect{x}_j=\vect{y_j}}=g'(\vect{y}; \vect{q}), \;\vect{y}_j+\vect{q}_j \in \mathcal{Z}_j $.
\end{asu}

Assumptions \ref{1A} and \ref{1B} guarantee that the proximal upper-bound function $h$ must be a global upper-bound function of the objective function $g$. Furthermore, Assumption \ref{1C} guarantee that $h(\vect{x}_j,\vect{y})$ takes steps proportional to the negative of the gradient of the objective function $g(\vect{x}_j,\vect{y}_{-j})$ in the direction $\vect{q}$, i.e., the existence of first-order derivative behavior.

For ease of presentation, we use the following  proximal upper-bound, where the upper-bound is constructed through adding quadratic penalization to the objective function: 
\begin{equation}
\begin{aligned}
h(\vect{x_j},\vect{y})=g(\vect{x}_j,\vect{y}_{-j})+\frac{ \varrho}{2} ({\vect{x}_j-\vect{y}_j })^2,
\end{aligned}
\label{eq:approximation_BCD}
\end{equation}
where $\varrho$ is a positive penalty parameter. 

At each iteration $t$, the BSUM solves the proximal upper-bound function via the following update:
\begin{equation}
\begin{cases}
&\vect{x}^t_j \in \underset{\vect{x}_j \in \mathcal{Z}_j}{\text{argmin}}\; h({\vect{x}_j, \vect{x}_j^{t-1}}),\; \forall j \in \mathcal{J}^t,\\
& \vect{x}^{t}_k=\vect{x}^{t-1}_k, \; \forall k \notin  \mathcal{J}^t.
\end{cases}
\label{eq:optimization_bsum_update}
\end{equation} 
There are many selection rules that can be used for selecting each coordinate $j \in \mathcal{J}^t$ at each iteration $t$, such as Cyclic, Gauss-Southwell, and Randomize \cite{hong2016unified}. The complete structure of the BSUM algorithm is described in Algorithm \ref{algo:BSUM}.
\begin{algorithm}[H]
	\caption{: BSUM algorithm in its standard form \cite{hong2016unified}}
	\label{algo:BSUM}
	\begin{algorithmic}[1]
		\STATE{\textbf{Input:} $\vect{x}$; }
		\STATE{\textbf{Output:} $\vect{x}^*$};
		\STATE {Initialize $t=0$, $\epsilon >0$;}
		\STATE {Find a feasible point $\vect{x}^0 \in \mathcal{Z} $;}
		\STATE{\textbf{Repeat};}
		\STATE{Choose index set $\mathcal{J}^t$;}
		\STATE{Let ${\vect{x}^t_j \in \text{argmin}}\; h({\vect{x}_j, \vect{x}^{t-1}_{-j}}),\;\forall j \in \mathcal{J}^t$;}
		\STATE{Set $\vect{x}^{t}_k=\vect{x}^{t-1}_k, \;\forall k \notin  \mathcal{J}^t$;}
		\STATE{$t=t+1$;}
		\STATE{\textbf{Until } $\lVert\frac{h_j^{(t)}-h_j^{(t+1)}}{h_j^{(t)}}\rVert  \leq \epsilon $;}\\
		\STATE{Then, consider  $\vect{x}^*=\vect{x}_j^{(t+1)}$ as solution.}
	\end{algorithmic}
\end{algorithm}

Algorithm \ref{algo:BSUM} (BSUM) can be considered as a generalized form of BCD that optimizes block by block the upper-bound function of the original objective function. BSUM can be used for solving separable smooth or non-smooth convex optimization problems that have linear coupling constraints. To solve the family of such problems, the BSUM updates each block of variables iteratively through minimizing the proximal upper-bound function until it converges to both a coordinate-wise minimum and a stationary solution. We consider the stationary solution to be a coordinate-wise minimum, when a block of variables reaches the minimum  point $\vect{x}^*=\vect{x}_j^{(t+1)}$. In other words, at stationary points, the entire vector of points cannot find a better minimum direction \cite{hong2016unified, hong2014block}. Based on \cite{hong2016unified} and \cite{hong2017iteration}, we can make the following remark:

\begin{remark}[Convergence] BSUM algorithm  takes $\mathcal{O}\left(\log(1/\epsilon)\right)$ to converge to an $\epsilon$-optimal solution, which is sub-linear convergence.
\end{remark}

The $\epsilon$-optimal solution $\vect{x}^\epsilon_j \in \mathcal{Z}_j$ is defined as $\vect{x}^\epsilon_j \in \{\vect{x}_j|\vect{x}_j \in \mathcal{Z}_j, \;h(\vect{x}_j,\vect{x}^t, \vect{y}^t)-h(\vect{x}^*_j,\vect{x}^t, \vect{y}^t)\}\leq \epsilon$, where $h(\vect{x}^*_j,\vect{x}^t, \vect{y}^t)$ is the optimal value of $h(\vect{x}_j,\vect{y})$ with respect  to $\vect{x}_j$.

\subsubsection{Distributed Optimization Control Algorithm}
\label{subsubsec:BSUM_4C} In our optimization problem in (\ref{eq:transformed_equation}) is difficult to solve due to the presence of decision variables that need to be used in different locations, and updating these variables one
at a time is impractical. Therefore, we consider BSUM  as a suitable candidate method for solving it in a distributed way by focusing on solving per-block subproblems. In order to apply BSUM in our distributed optimization control model, we define $\mathcal{X}\triangleq\{\vect{x}:\sum_{m \in \mathcal{M}}\sum_{k\in \mathcal{K}_m} x_k^m=1, x_k^m  \in [0,1]\}$,  $\mathcal{Y}\triangleq\{\vect{y}:\sum_{m \in \mathcal{M}}\sum_{k\in \mathcal{K}_m} y_k^{k\rightarrow m} + y^{m\rightarrow n}_k +y^{m\rightarrow DC}_k=1, y_k^{k\rightarrow m} , y^{m\rightarrow n}_k , y^{m\rightarrow DC}_k  \in [ 0, 1]\}$, and $ \mathcal{W}\triangleq \{\vect{w}: \sum_{m \in \mathcal{M}}\sum_{k\in \mathcal{K}_m} w_m^k  + w_n^k + w_{DC}^k= 1 , w_m^k, w_n^k , w_{DC}^k  \in [0,1]\}$ as the feasible sets of  $\vect{x}$,  $\vect{y}$, and $\vect{w}$, respectively.

At each iteration $t$, $\forall j \in \mathcal{J}^t$, we define the proximal upper-bound function $B_j$, which is convex and the proximal upper-bound of the objective function defined in (\ref{eq:transformed_equation}). In order to guarantee that the proximal upper-bound function $B_j$ is convex, we add to the objective function in (\ref{eq:transformed_equation}) a quadratic penalization, as follows:

\begin{equation}
\begin{aligned}
\mathcal{B}_j({\vect{x}_j,\vect{x}^{(t)}, \vect{y}^{(t)}, \vect{w}^{(t)}})\defeq \mathcal{B}{(\vect{x}_j,\vect{\tilde{x}}, \vect{\tilde{y}}, \vect{\tilde{w}}}) + \frac{ \varrho_j}{2} \lVert(\vect{x}_j- \vect{\tilde{x}})\rVert^2. 
\end{aligned}
\label{eq:optimization_bsum1}
\end{equation}

(\ref{eq:optimization_bsum1}) is the proximal upper-bound function of  (\ref{eq:transformed_equation}), and it can be applied to other vectors of variables $\vect{y}_j$ and $\vect{w}_j$, respectively, where $\varrho_t>0$ is the positive penalty parameter. Furthermore, the proximal upper-bound function in (\ref{eq:optimization_bsum1}) is a convex optimization problem due to its quadratic term $\frac{ \varrho_j}{2} \lVert(\vect{x}_j- \vect{\tilde{x}})\rVert^2$. In other words, with respect to  $\vect{x}_j$, $\vect{y}_j$, and $\vect{w}_j$, it has minimizers vector  $\vect{\tilde{x}}$, $\vect{\tilde{y}}$, and $\vect{\tilde{w}}$ at each iteration $t$, which are considered to be the solution of the previous step ($t-1$). At each iteration $t+1$, the solution is updated by solving the following optimization problems:
\begin{equation}
\begin{aligned}
\vect{x}_j^{(t+1)}\in \underset{ \vect{x}_j \in \mathcal{X}}{\text{min}}\; \mathcal{B}_j (\vect{x}_j,\vect{x}^{(t)}, \vect{y}^{(t)}, \vect{w}^{(t)}),
\end{aligned}
\label{eq:optimization20}
\end{equation} 
\begin{equation}
\begin{aligned}
\vect{y}_j^{(t+1)}\in \underset{ \vect{y}_j \in \mathcal{Y}}{\text{min}}\; \mathcal{B}_j (\vect{y}_j,\vect{y}^{(t)}, \vect{x}^{(t+1)}, \vect{w}^{(t)}),
\end{aligned}
\label{eq:optimization21}
\end{equation}
\begin{equation}
\begin{aligned}
\vect{w}_j^{(t+1)}\in \underset{ \vect{w}_j \in \mathcal{W}}{\text{min}}\; \mathcal{B}_j (\vect{w}_j,\vect{w}^{(t)}, \vect{x}^{(t+1)}, \vect{y}^{(t+1)}).
\end{aligned}
\label{eq:optimization22}
\end{equation}

Furthermore, (\ref{eq:optimization20}), (\ref{eq:optimization21}), and (\ref{eq:optimization22}) can be solved through the use of our proposed distributed optimization control presented in  Algorithm \ref{algo:onlienalgorithm} for 4C, which is a modified version of the standard BSUM (Algorithm \ref{algo:BSUM}). For solving (\ref{eq:optimization20}), (\ref{eq:optimization21}), and (\ref{eq:optimization22}), we relax the  vectors of variables $\vect{x}_j$, $\vect{y}_j$, and $\vect{w}_j$ taking values in the closed interval between $0$ and $1$. Then, we use  a threshold rounding technique described in \cite{feige2016oblivious} in Algorithm \ref{algo:onlienalgorithm} to enforce the relaxed $\vect{x}_j$, $\vect{y}_j$, and $\vect{w}_j$ to be vectors of binary variables.

As an example, in the rounding technique, for $x_k^{m*}  \in \vect{x}_j^{(t+1)}$, $x_k^{m*} \geq \theta$, where $\theta \in (0,1)$ is a positive rounding threshold, we set $x_k^{m*}$  as follows: 

\begin{equation}
\label{eq:b_rounding}
\setlength{\jot}{10pt}
x_k^{m*} =
\begin{cases}
1,\; \text{if $x_k^{m*} \geq \theta$},\\
0, \;\text{otherwise.}
\end{cases}
\end{equation}

The above rounding technique can be applied to other vectors of variables $\vect{y}_j$ and $\vect{w}_j$, respectively. However, the binary solution obtained from the rounding technique may violate communication, computational, and caching resource constraints. Therefore, as described in \cite{zhang2017network}, to overcome this issue after rounding, we solve the  problem (\ref{eq:optimization_bsum1}) in the form of $\mathcal{B}_j + \xi \Delta$, where constraints $(\ref{first:a})$, $(\ref{first:b})$, and $(\ref{first:c})$ are modified as follows: 
\begin{equation}
\sum_{k\in \mathcal{K}_m}x_k^ma^m_{k}\leq 1 + \Delta_a, \;  \forall m \in \mathcal{M},\label{first:a_m}
\end{equation}	
\begin{equation}
\label{first:b_m}
\sum_{k\in \mathcal{K}_m}x_k^mp_{km}y_k^{k\rightarrow m}\leq P_m + \Delta_p,\forall  m \in \mathcal{M},
\end{equation}	
\begin{equation}
\label{first:c_m}
x_k^m (\sum_{k\in \mathcal{K}_m} y^{k \rightarrow m} + \sum_{n\neq m\in \mathcal{M}}\sum_{k\in \mathcal{K}_n}  y^{n\rightarrow m}_k) w_m^k s(d_k)\leq C_m + \Delta_m,
\end{equation}
where $\Delta_a$ is the maximum violation of communication resources constraint, $\Delta_p$ is the maximum violation of computational resources constraint, $\Delta_m$ is the maximum violation of caching resources constraint, $\Delta= \Delta_a + \Delta_p + \Delta_m$, and $\xi$ is the weight of $\Delta$. Moreover, $\Delta_a$,  $\Delta_p$,  and  $\Delta_m$ are given by:
\begin{equation}
\label{eq:rounding-10}
\Delta_a =\max \{ 0,\sum_{k\in \mathcal{K}_m}x_k^ma^m_{k} - 1 \},\;\forall  m \in \mathcal{M},
\end{equation}
\begin{equation}
\label{eq:rounding-11}
\Delta_p = \max \{0, \sum_{k\in \mathcal{K}_m}x_k^mp_{km}y_k^{k\rightarrow m} - P_m  \},\;\forall  m \in \mathcal{M},
\end{equation}
\begin{multline}
\label{eq:rounding-22}
\Delta_m =\max \{0, x_k^m (\sum_{k\in \mathcal{K}_m} y^{k \rightarrow m} + \sum_{n\neq m\in \mathcal{M}}\sum_{k\in \mathcal{K}_n}  y^{n\rightarrow m}_k) w_m^k s(d_k)\\- C_m  \}.
\end{multline}
Furthermore, if there are no violations of communication, computational, and caching resources constraints ($\Delta_a=0$, $\Delta_p=0$, and $\Delta_m=0$), the feasible solution of (\ref{eq:optimization_bsum1}) is obtained.

\textcolor{black}{Given problem $\mathcal{B}_j$  and its rounded problem $\mathcal{B}_j + \xi \Delta$, a most important measurement of the quality of rounding technique is the integrality gap, which measure the ratio between the feasible solutions of  $\mathcal{B}_j$ and  $\mathcal{B}_j + \xi \Delta$. Therefore, based on definition and proof of integrality gap in  \cite{feige2016oblivious}, we can make the following definition:}  
\begin{definition}[Integrality gap] Given problem $\mathcal{B}_j$ (\ref{eq:optimization_bsum1})  and its rounded problem $\mathcal{B}_j + \xi \Delta$, the integrality gap is given by:\end{definition}
\textcolor{black}{\begin{equation}
	\beta=\underset{\vect{x}, \vect{y}, \vect{w}}{\text{min}}\ \ \frac{\mathcal{B}_j}{\mathcal{B}_j + \xi \Delta}, 
	\end{equation}where the solution of $\mathcal{B}_j$ is obtained through relaxation of variables $\vect{x}_j$, $\vect{y}_j$, and $\vect{w}_j$, while the solution of $\mathcal{B}_j + \xi \Delta$ is obtained after rounding the relaxed variables. We consider that the best rounding is achieved, when $\beta$ ($\beta \leq 1$) is closer to $1$ \cite{feige2016oblivious}.  In other words, $\beta=1$, when $\Delta_a=0$, $\Delta_p=0$, and $\Delta_m=0$.}


In Algorithm \ref{algo:onlienalgorithm} for 4C, each  user device $k \in \mathcal{K}$ chooses the offloading  decision $x_k^m$. If $x_k^m=1$, the user sends its demands to the nearest BS. For each demand $T_{k}$ received, the BS checks its RAT for its own and collaboration space resource availabilities. Algorithm \ref{algo:onlienalgorithm} starts by initializing $t=0$, and setting $\epsilon$ equal to a small positive number, where $\epsilon$ is used to guarantee the  $\epsilon$-optimal solution defined in \cite{hong2016unified}.  Algorithm \ref{algo:onlienalgorithm} then finds the initial feasible points ($\vect{x}^{(0)}$, $\vect{y}^{(0)}$, $\vect{w}^{(0)}$). Subsequently, our algorithm starts an iterative process and chooses the index set. At each iteration $t+1$, the solution is updated by solving the optimization problems (\ref{eq:optimization20}), (\ref{eq:optimization21}), and (\ref{eq:optimization22})  until $\frac{\mathcal{B}_j^{(t)}-\mathcal{B}_j^{(t+1)}}{\mathcal{B}_j^{(t)}} \leq \epsilon $. Algorithm \ref{algo:onlienalgorithm} generates a binary solution of $\vect{x}_j^{(t+1)}$, $\vect{y}_j^{(t+1)}$, and $\vect{w}_j^{(t+1)}$  via the
rounding technique (\ref{eq:b_rounding}), solves $\mathcal{B}_j + \xi \Delta$, and calculates $\beta$ for obtaining $\vect{c}$,  $\vect{p}$, and $\vect{R}$. Furthermore, $\vect{x}^*=\vect{x}_j^{(t+1)}$, $\vect{y}^*=\vect{y}_j^{(t+1)}$, and $\vect{w}^*=\vect{w}_j^{(t+1)}$ are considered to be stationary solution that satisfies coordinate-wise minimum. Finally, Algorithm 3 updates its RAT and sends the RAT update in its collaboration space.
\begin{algorithm}[t]	
	\caption{: Distributed optimization control algorithm (BSUM-based) for 4C in big data MEC}
	\label{algo:onlienalgorithm}
	\begin{algorithmic}[1]
		\STATE{\textbf{Input:} $\vect{T}$: A vector of demands;  $B_m$, $P_m$, and $C_m$: communication, computational and caching resources; }
		\STATE{\textbf{Output:} $\vect{x}^*, \;\vect{y}^*, \;\vect{w}^*$,  $\vect{c}$ : A vector of cache allocation, $\vect{p}$: A vector of computation allocation, and $\vect{R}$: A vector of communication resources allocation};
		\STATE { Each user device $k \in \mathcal{K}$ chooses the offloading\\ decision $x_k^m$;}
		\STATE{If $x_k^m=1$, user device $k \in \mathcal{K}$ sends its demand $T_{k}$ to BS $m \in \mathcal{M}$;}
		\STATE {For each $T_{k}$ received at BS $m \in \mathcal{M}$, check RAT update;}
		\STATE {Initialize $t=0$, $\epsilon >0$;}
		\STATE {Find initial feasible points ($\vect{x}^{(0)}$,  $\vect{y}^{(0)}$, $\vect{w}^{(0)}$);}
		\REPEAT
		\STATE{Choose index set $\mathcal{J}^t$;}
		\STATE{Let $\vect{x}_j^{(t+1)}\in \underset{ \vect{x}_j \in \mathcal{X}}{\text{min}}\; \mathcal{B}_j (\vect{x}_j,\vect{x}^{(t)},\vect{y}^{(t)}, \vect{w}^{(t)})$;}
		\STATE{Set $\vect{x}_k^{t+1}=\vect{x}_k^{t}, \forall k \notin  \mathcal{J}^t$;}
		\STATE{Go to Step $4$,  find  $\vect{y}_j^{(t+1)}$, $\vect{w}_j^{(t+1)}$ by solving (\ref{eq:optimization21}) and (\ref{eq:optimization22});}	
		\STATE{$t=t+1$;}	
		\UNTIL{ $\lVert\frac{\mathcal{B}_j^{(t)}-\mathcal{B}_j^{(t+1)}}{\mathcal{B}_j^{(t)}}\rVert  \leq \epsilon $;}
		\STATE{Generate a binary solution of $\vect{x}_j^{(t+1)}$, $\vect{y}_j^{(t+1)}$, and $\vect{w}_j^{(t+1)}$  via the
			rounding technique (\ref{eq:b_rounding}), solve $\mathcal{B}_j + \xi \Delta$, and calculate $\beta$ for obtaining $\vect{c}$,  $\vect{p}$, and $\vect{R}$;}
		\STATE{Then, consider $\vect{x}^*=\vect{x}_j^{(t+1)}$, $\vect{y}^*=\vect{y}_j^{(t+1)}$, and $\vect{w}^*=\vect{w}_j^{(t+1)}$ as a solution;}
		\STATE{Update RAT, and send RAT update in collaboration space.}
	\end{algorithmic}
\end{algorithm}


The difference between the BSUM (Algorithm  \ref{algo:BSUM}) in its standard form and the BSUM for 4C in big data MEC (Algorithm \ref{algo:onlienalgorithm}) resides in their implementations, where BSUM Algorithm in its standard form is  based on distributed control. On the other hand, Algorithm \ref{algo:onlienalgorithm} is based on both the hierarchical and distributed control models defined in \cite{molzahn2017survey}. In the  hierarchical control model, edge devices decide on $\vect{x}$ first. Then, each MEC server $m$ acts as a controller for the users' offloaded tasks and, thus, it solves (\ref{eq:optimization20}), (\ref{eq:optimization21}), and (\ref{eq:optimization22}).

In the distributed control model, each MEC server exchanges small information with other MEC servers in order to maintain the resource allocation within a tight range of available computational resources $P$ and  caching resources $C$. However, in a collaboration space, there is no centralized controller that controls all MEC servers. This distributed control is modeled as a dynamic feedback control model described in \cite{farivar2015local}, where the RAT update at each MEC server acts as feedback with state $(\vect{x}^{(t)},\;\vect{y}^{(t)},\; \vect{w}^{(t)})$ at iteration $t$,  which is used to determine the new state $(\vect{x}^{(t+1)},\;\vect{y}^{(t+1)}, \;\vect{w}^{(t+1)})$ at the next iteration $t+1$. Furthermore, the optimal value ($\vect{x}^*_j, \;\vect{y}^*_j, \;\vect{w}^*_j$) is considered to be network equilibrium or a stability point, which is the stationary solution that satisfies a coordinate-wise minimum. 

\section{Simulation Results and Analysis}
\label{sec:PE}
In this section, we present the performance evaluation of the proposed joint 4C in big data MEC, where the Python language \cite{van2007python} is used for numerical analysis.

\subsection{Simulation Setup}
For forming collaboration spaces, we use the Sitefinder dataset from Edinburgh DataShare \cite{boswarva2017sitefinder}. In this dataset, we randomly select one MNO, which has $12777$ BSs, through use of the OKM-CS algorithm for unsupervised machine learning, where we group these BSs into $1000$ collaboration spaces. Therefore, based on the BS locations and their proximities, the number of BSs in   one collaboration space is in the range from $1$ BS to $203$ BSs. Among $1000$ collaboration spaces, we randomly select one collaboration space, which has $12$ BSs, and we associate each BS with $1$ MEC server. Furthermore, we consider the initial number of users to be $K=50$ at each BS. where each user sends one task at each time slot. The path loss factor is set to $4$ and the transmission power is set to $\rho_k=27.0$ dBm \cite{zhou2017resource}, while the channel bandwidth is set to be in the range from  $B_m=25$ MHz to $B_m=32$ MHz \cite{boswarva2017sitefinder}. Furthermore, we consider the bandwidth between each pair of BSs to be randomly selected in the range from $\Gamma^n_m=20$ MHz to $\Gamma^n_m=25$ MHz, while the bandwidth between each BS and DC is selected in the range from $\Omega^{DC}_m=50$ to $\Omega^{DC}_m=120$ Mbps. The cache storage of each MEC server $m$ is in the range from $100$ to $500$ TB, while computation resources are in the range from $2$ GHz to $2.5$ GHz\cite{mao2017stochastic}.

For task $ T_{k}$ of a given user $k$, the size of data $s(d_k)$ is randomly generated in the range from $2$ to $7$ $GB$, while the task computation deadline $\tilde{\tau}_{k}$ is randomly generated  in  the range from $\tilde{\tau}_{k}=0.02$ second to $\tilde{\tau}_{k}=12$ seconds. The workload $z_k$ of each user device $k$ is randomly generated and uniformly distributed in the range from $z_k= 452.5$ cycles/bit to $z_k=737.5$ cycles/bit \cite{mao2017stochastic}. For each user device, the computation resource is in range from $0.5$ GHz to $1.0$ GHz \cite{chen2016efficient}. We consider that each end-user device has a CPU peak bandwidth of $16$-bit values per cycle, while each MEC server has a CPU peak bandwidth of 64-bit values per cycle.

At each time slot, we use $50$ different contents, where the total number of requests for contents ranges from $\lambda^{d_k}_m=578$ to $\lambda^{d_k}_m=3200$. The demand and popularity of the content follow Zipf distributions as described in \cite{newman2005power, ndikumana2017network}.

\subsection{Performance Metrics}
\subsubsection{Throughput}
\textcolor{black}{For effective resource utilization, we evaluate the network and computation throughputs of the proposed algorithms}. We define the network throughput as a measurement of how many units of information that a network can handle for a given period of time \cite{ndikumana2017novel, ndikumana2015network}, while computation throughput is defined as a measurement of how many units of tasks that a given MEC server can compute for a given period of time.  Here, the network throughput is measured in terms of Mbps, while the computation throughput is measured in terms of million instructions per second (MIPS).

\subsubsection{Delay}
In a collaboration space, each task $T_ {k} $ offloaded by the user device ends its journey at the server which has resources that can fulfill user demand. Then, the MEC server computes, caches, and returns the output of the computation to the user. Therefore, we consider the total delay as the time period between offloading task $T_ {k} $ and receiving the corresponding computation output.
\subsubsection{Cache Hit Ratio and Bandwidth-saving}
We also evaluate the number of cache hits and misses. A cache hit,  denoted $h^{d_k}_m \in \{  0, 1\}$, occurs when the requested content $d_k$  is retrieved from the cache storage available in a collaboration space at any BS $m$. Cache hit contributes to  bandwidth saving defined in (\ref{eq:caching_reward}) as it reduces the data exchange between the collaboration space and the DC. On the other hand, a cache miss occurs when the requested content $d_k$ is not available in any cache storage in the collaboration space. 
The probability of a cache hit for content $d_k$ is expressed as follows:
\begin{equation}
\begin{aligned}
P_{d_k} = \frac{\sum_{k\in \mathcal{K}}\sum_{m\in \mathcal{M}} h^{d_k}_m} {\sum_{k\in \mathcal{K}}\sum_{m\in \mathcal{M}} (h^{d_k}_m + (1 - h^{d_k}_m))},
\end{aligned}
\label{eq:zipf-distribution}
\end{equation}
where  $\sum_{k\in \mathcal{K}}\sum_{m\in \mathcal{M}} h^{d_k}_m$ is the total number of cache hits, and $\sum_{k\in \mathcal{K}}\sum_{m\in \mathcal{M}} (h^{d_k}_m + (1 - h^{d_k}_m))$ is the total number of cache hits plus the total number of cache misses.

\subsection{Simulation Results}
\begin{figure}[t]
	\centering
	\begin{minipage}{0.45\textwidth}
		\centering
		\includegraphics[width=1.0\columnwidth]{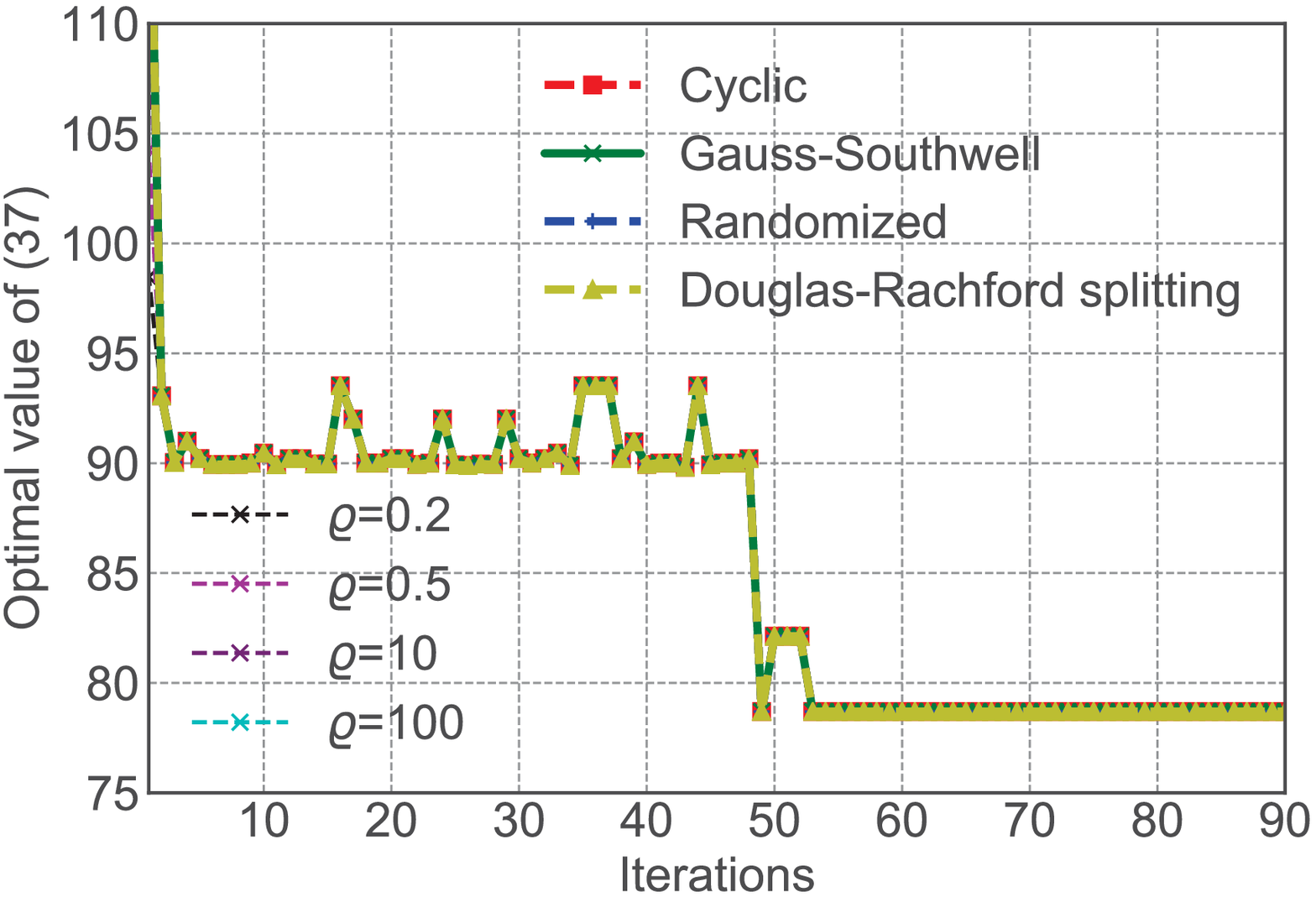}
		\caption{Optimal value of (\ref{eq:optimization_bsum1}) with different coordinate selection rules (without rounding).}
		\label{fig:Optimization_comparization1}
	\end{minipage}	
	\begin{minipage}{0.45\textwidth}
		\centering
		\includegraphics[width=1.0\columnwidth]{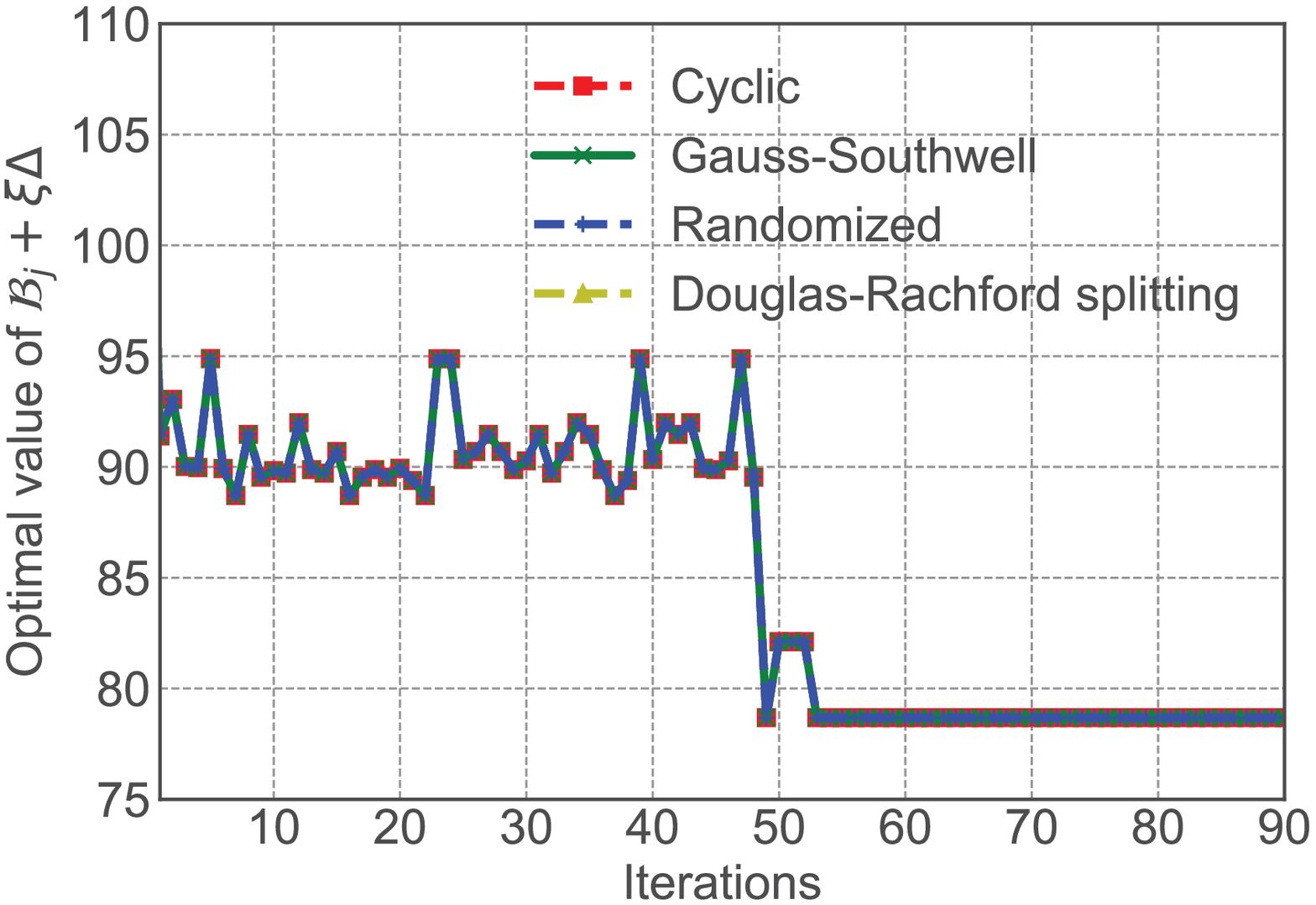}
		\caption{Optimal value of $\mathcal{B}_j + \xi \Delta$ with different coordinate selection rules (after rounding).}
		\label{fig:Optimization_comparization2}
	\end{minipage}
\end{figure}
\begin{figure}[t]
	\centering
	\begin{minipage}{0.45\textwidth}
		\centering
		\includegraphics[width=1.0\columnwidth]{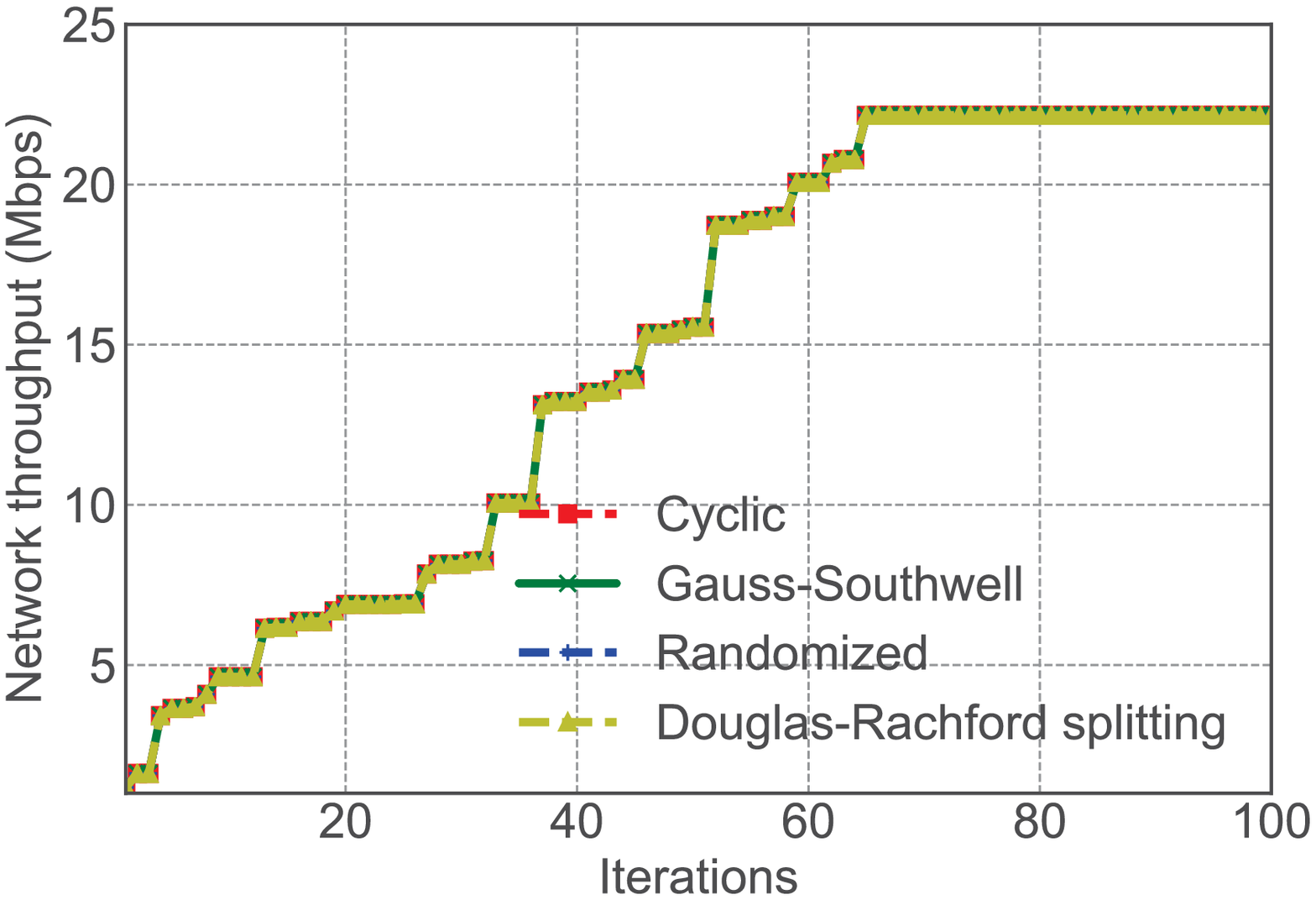}
		\caption{Network throughput within collaboration space. }
		\label{fig:comunication_allocation}
	\end{minipage}
	\begin{minipage}{0.45\textwidth}
		\centering
		\includegraphics[width=1.0\columnwidth]{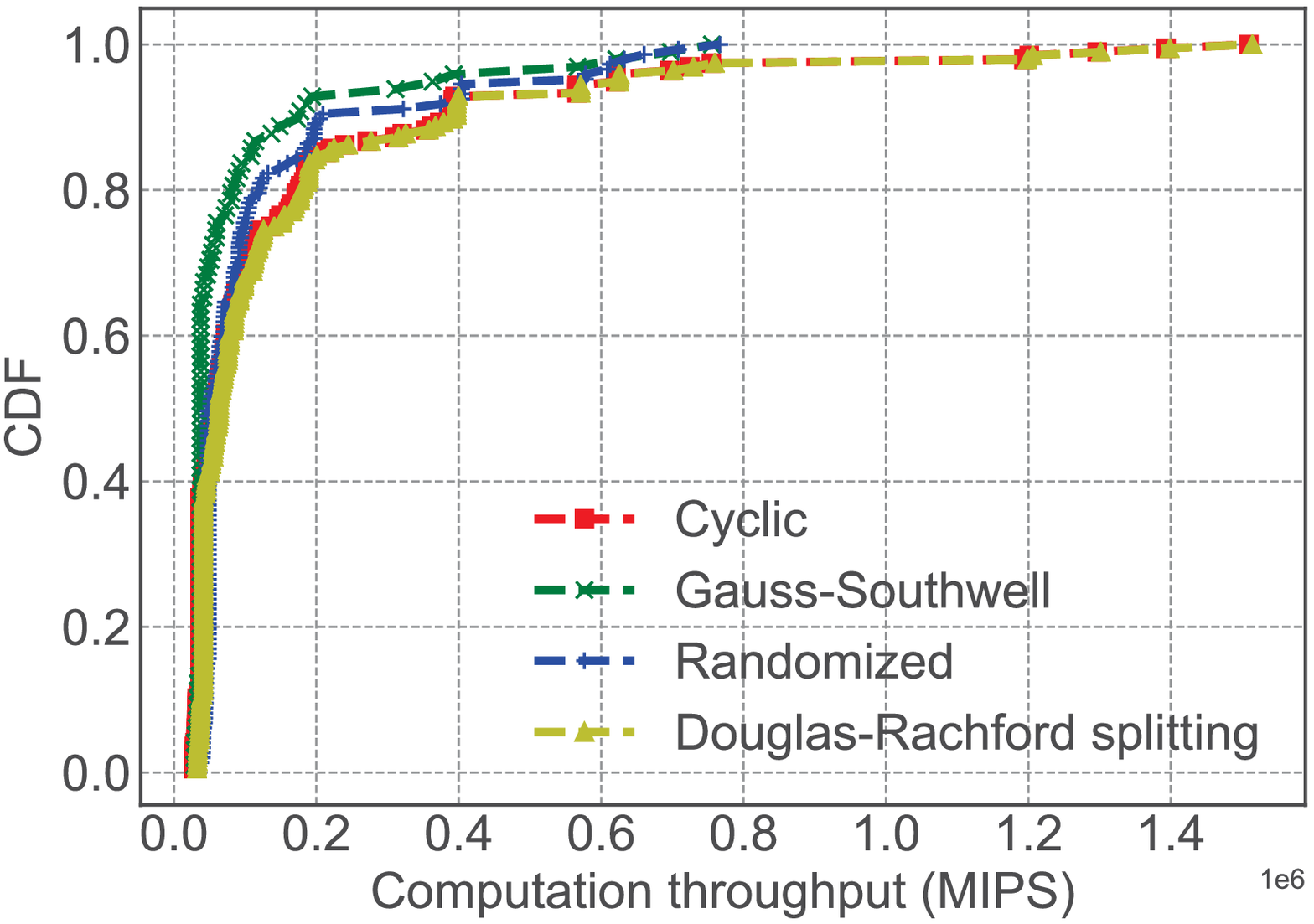}
		\caption{CDF of computation throughput.}
		\label{fig:computation_allocation}
	\end{minipage}	
\end{figure}

\begin{figure}[t]
	\centering
	\begin{minipage}{0.45\textwidth}
		\centering
		\includegraphics[width=1.0\columnwidth]{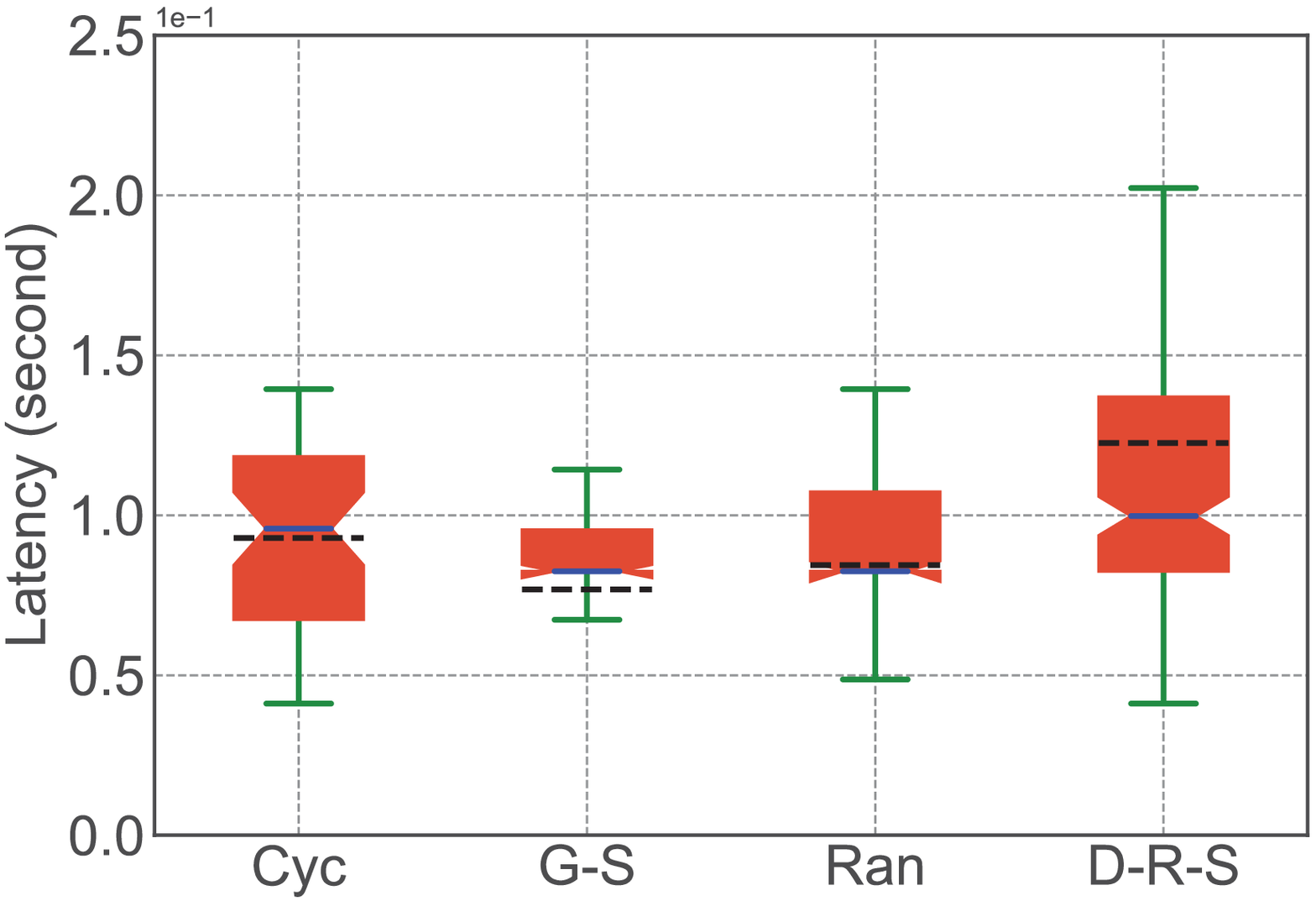}
		\caption{Total delay for offloading and computation.}
		\label{fig:latency}
	\end{minipage}	
	\begin{minipage}{0.45\textwidth}
		\centering
		\includegraphics[width=1.0\columnwidth]{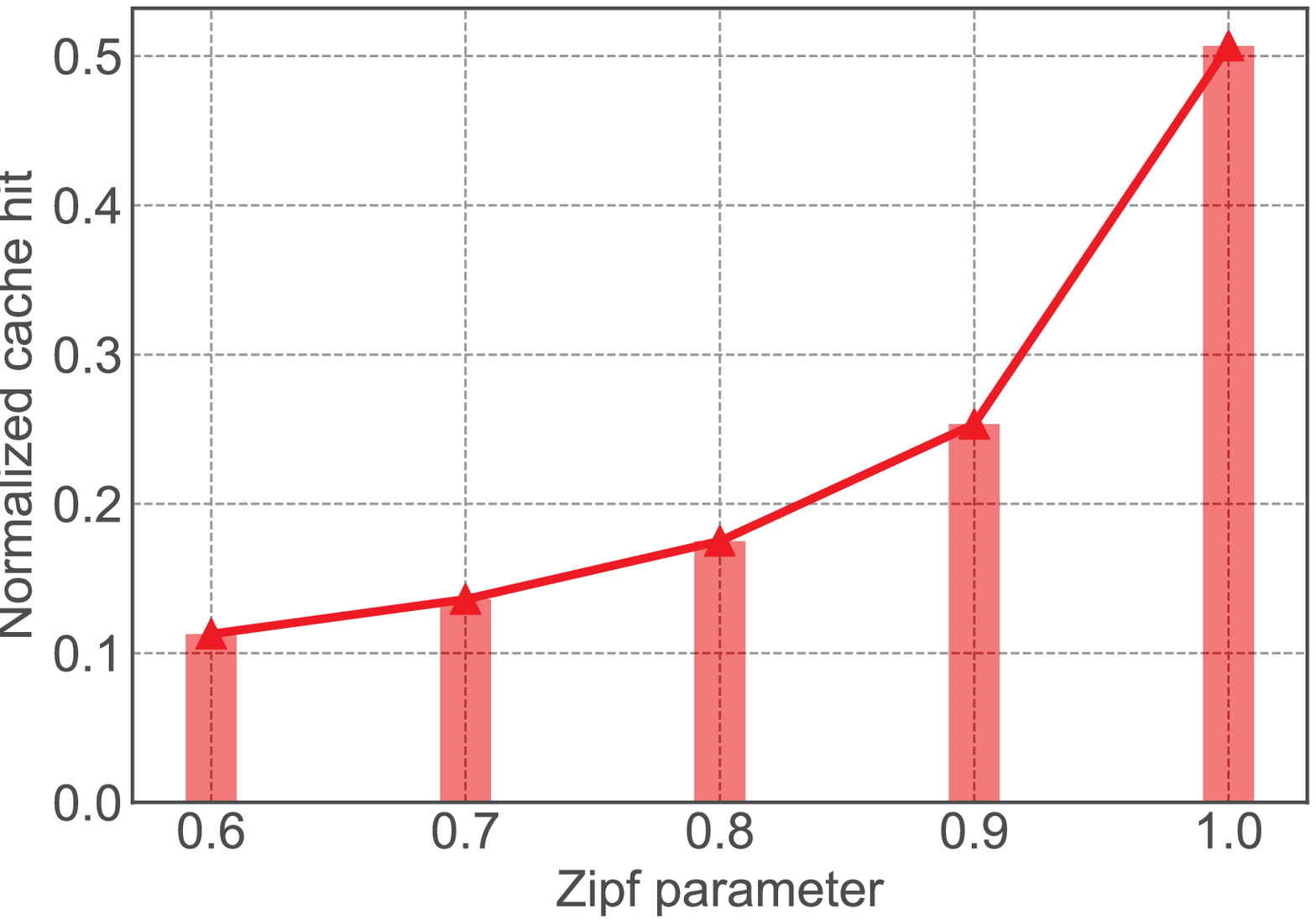}
		\caption{Normalized cache hits in collaboration space.}
		\label{fig:normalized_cache_hit1}
	\end{minipage}		
\end{figure}
\begin{figure}[t]
	\centering
	\begin{minipage}{0.45\textwidth}
		\centering
		\includegraphics[width=1.0\columnwidth]{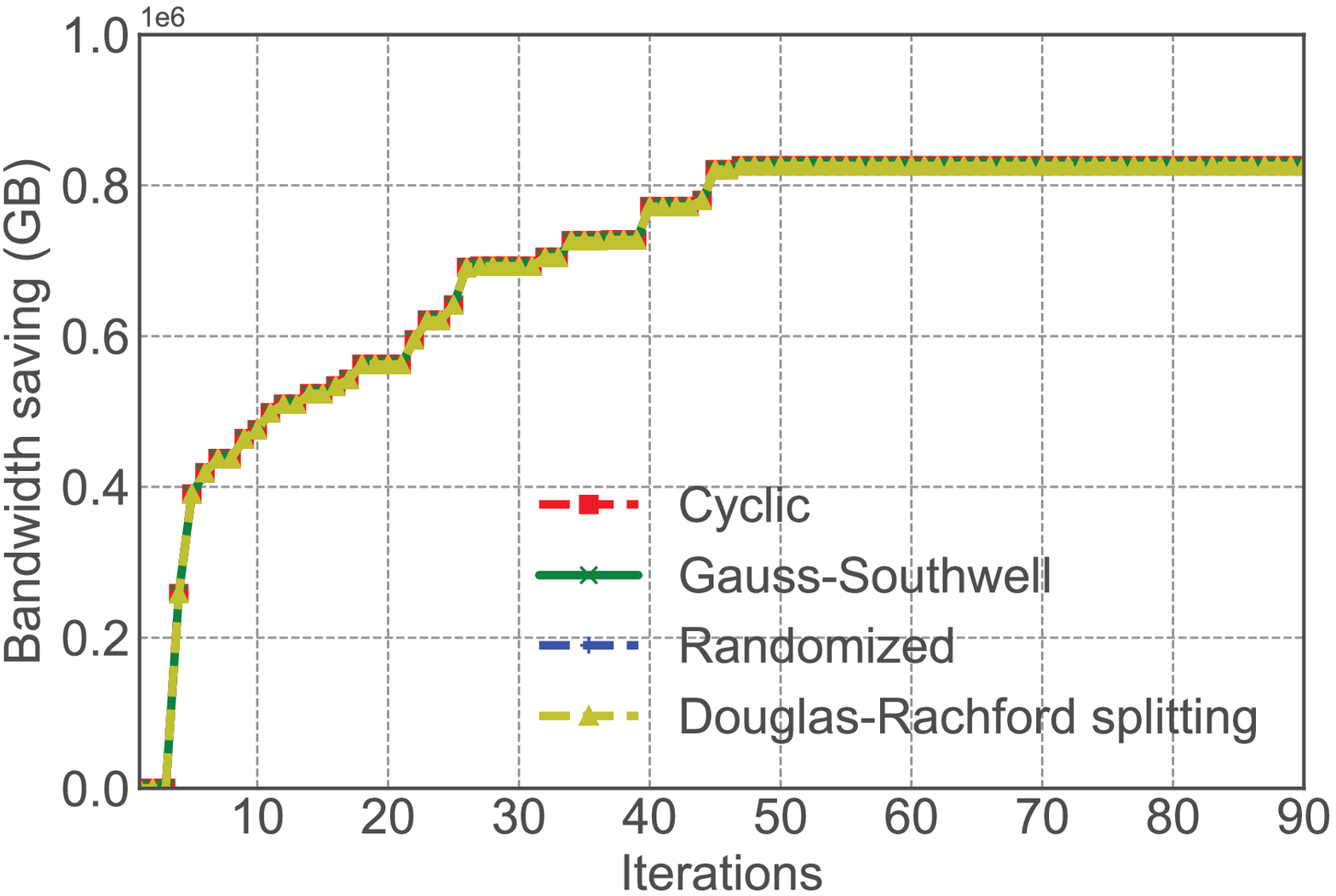}
		\caption{Bandwidth saving  due to caching.}
		\label{fig:Bandwidth-saving2}
	\end{minipage}
	\begin{minipage}{0.45\textwidth}
		\centering
		\includegraphics[width=1.0\columnwidth]{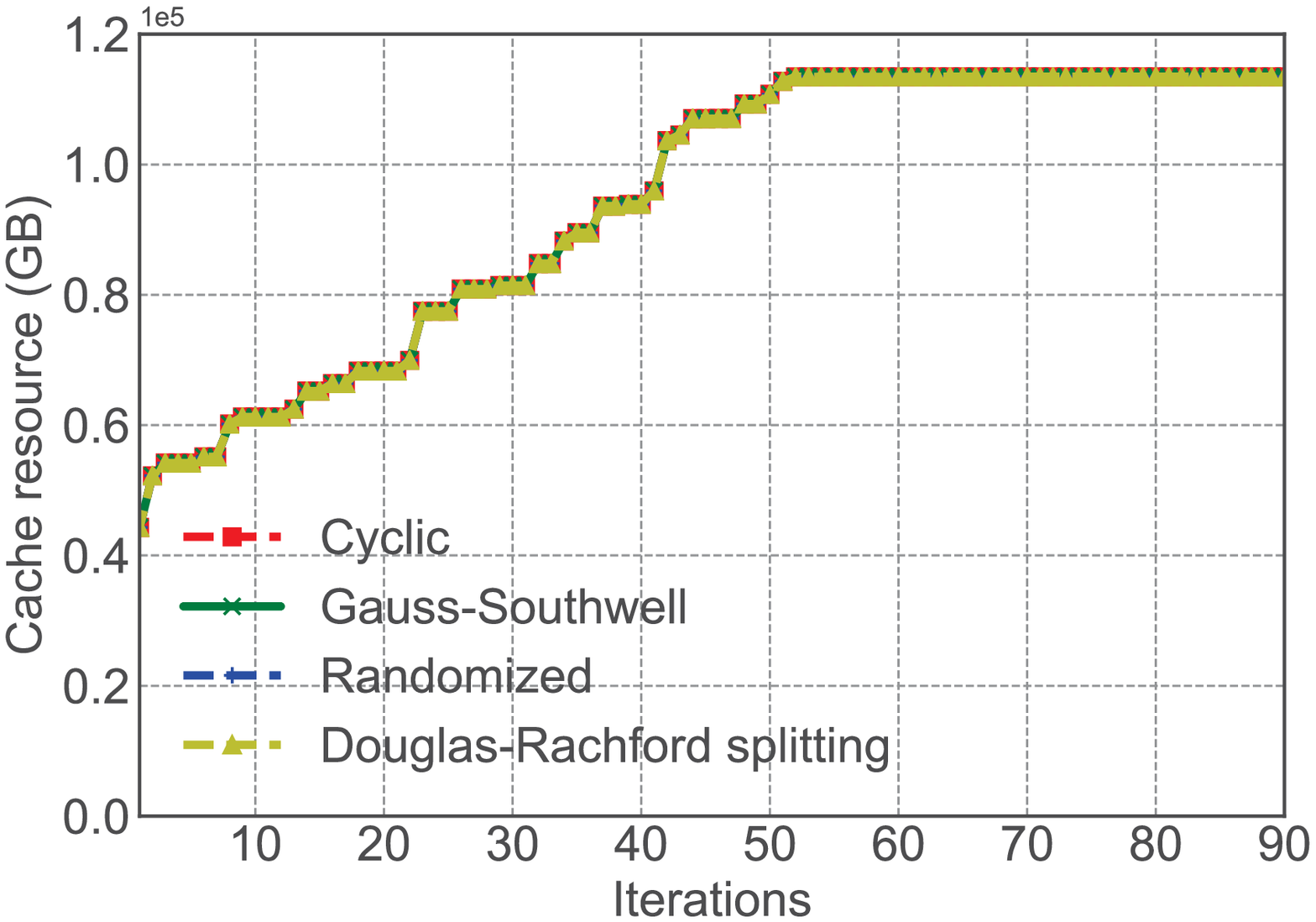}
		\caption{Utilization of MEC cache storages.}
		\label{fig:caching_allocation}
	\end{minipage}	
\end{figure}

Fig. \ref{fig:Optimization_comparization1} combines both delay viewed as cost and bandwidth saving in one optimization problem in (\ref{eq:optimization_bsum1}). We solve the proximal upper-bound problem through  use of  distributed optimization control algorithm for 4C (Algorithm \ref{algo:onlienalgorithm}) and CVXPY \cite{diamond2016cvxpy} (a Python-embedded modeling language for solving convex optimization problems). Furthermore, we compare the solution of our distributed optimization control algorithm with the solution computed via Douglas-Rachford splitting \cite{themelis2017douglas} without applying a rounding technique. Thus, our formulated problem in (\ref{eq:optimization_bsum1}) is decomposable. The Douglas-Rachford splitting method is used to decompose our problem into small subproblems, and address each subproblem separately.

Fig. \ref{fig:Optimization_comparization1} shows the convergence of our optimization problem without rounding. In  this figure, we use the Douglas-Rachford splitting method \cite{themelis2017douglas} and our distributed control algorithm (Algorithm \ref{algo:onlienalgorithm}) for solving (\ref{eq:optimization_bsum1}). In our distributed control algorithm, for choosing indexes in  $(\ref{eq:optimization_bsum1})$, we  use  three  coordinate selection rules: Cyclic, Gauss-Southwell, and Randomized \cite{hong2016unified}. Furthermore, for the quadratic term in ($\ref{eq:optimization_bsum1}$), we adjust the positive penalty parameter $\varrho_j$ within the range $0.2$ to $100$. From this figure, we can see that the performance of our distributed control algorithm and Douglas-Rachford splitting method is almost nearly the same. Therefore, as of iteration $53$, the proximal upper-bound problem in $(\ref{eq:optimization_bsum1})$ converges to both a coordinate-wise minimum and a stationary point, which is considered as a solution of $(\ref{eq:optimization_bsum1})$. In other words, we consider this minimum point as an optimal value and equilibrium/stability point of $\mathcal{B}_j$ ($\ref{eq:optimization_bsum1}$).

In Fig. \ref{fig:Optimization_comparization2}, we apply the rounding technique to the results of Fig. \ref{fig:Optimization_comparization1} and solve $\mathcal{B}_j + \xi \Delta$, where we consider the positive rounding threshold to be $\theta=7$ and the weight of $\Delta$ to be $\xi=0.14285$. The simulation results in Fig.  \ref{fig:Optimization_comparization2} ensure that the relaxed $\vect{x}_j$, $\vect{y}_j$, and $\vect{w}_j$ to be vectors of binary variables, and the rounding technique does not violate the computational and caching resource constraints while solving $\mathcal{B}_j + \xi \Delta$. Furthermore, the difference between Fig. \ref{fig:Optimization_comparization1} and Fig. \ref{fig:Optimization_comparization2} resides in the sizes of the problems ($\mathcal{B}_j$ and $\mathcal{B}_j + \xi \Delta$) and the step sizes needed for reaching the minimum point. \textcolor{black}{ However, as of iteration 53, in both  Figs. \ref{fig:Optimization_comparization1} and \ref{fig:Optimization_comparization2},  both problems $\mathcal{B}_j$ and $\mathcal{B}_j + \xi \Delta$ converge to the same stability point. In other words, with and without applying rounding technique, $(\ref{eq:optimization_bsum1})$ converges to a minimum point that guarantees $\beta=1$ ( no violations of communication, computational, and caching resources constraints).}

\textcolor{black}{ In terms of network throughput,  Fig. \ref{fig:comunication_allocation} shows that the throughput increases up to $22.48$ Mbps. In this figure, the coordinate selection rules (Cyclic, Gauss-Southwell, Randomized) in our distributed optimization control algorithm and the Douglas-Rachford splitting method have almost the same performance.}

Fig. \ref{fig:computation_allocation} shows the cumulative distribution function (CDF) of the computational throughput. The simulation results show that the Cyclic selection rule in our distributed optimization control algorithm as well as the Douglas-Rachford splitting method require high computational resources, as the computational throughput for each MEC server can reach $1.55 \times 10^{6}$ MIPS. On the other hand, the Gauss-Southwell and Randomized selection rules use less computational resources, as the computational throughput for each MEC server can reach $0.78 \times 10^{6}$ MIPS. \textcolor{black}{ The advantage of the Gauss-Southwell selection rule compared to other coordinate selection rules lies in choosing the index. In the Gauss-Southwell selection rule, instead of choosing the index randomly or cyclically, at each iteration, an index that maximizes the utilization of the computational resource is chosen.}

We next examine the delay between offloading task $T_ {k} $ and receiving the corresponding computation output. Fig. \ref{fig:latency} shows the total delay experienced by user demand $ T_{k}$, where the solid blue lines represent the median and the dashed black lines represent the arithmetic mean.  In this figure, Cyc stands for Cyclic, G-S stands for Gauss-Southwell, Ran stands for Randomized, while D-R-S stands for Douglas-Rachford splitting. The results in this figure show that the mean of the delay varies from $\tilde{\tau}_{k}=0.077$ to $\tilde{\tau}_{k}=0.128$ seconds, which fulfills the task computation deadline described in the simulation setup. However, Cyclic and Douglas-Rachford splitting yield higher delay than others due to index selection (for Cyclic) and splitting (for Douglas-Rachford splitting), which require more time and computation resources. Furthermore, Douglas-Rachford splitting has a higher delay than BSUM coordinate selection rules.

Fig. \ref{fig:normalized_cache_hit1} shows the normalized cache hits, where cache hit ratio $P_{d_k}$ is computed from (\ref{eq:zipf-distribution}). From Fig. \ref{fig:normalized_cache_hit1}, we can see that the cache hit ratio increases with the Zipf exponent parameter $a$. When $a=1.0$, due to the increase in the number of demands for contents, many contents become popular, which results in a high cache hit ratio of $0.51\%$ of the total demands $\lambda^{d_k}_m$ from users. In the case of cache misses in collaboration space, the demands for contents need to be forwarded to the DC. Therefore, cache hits contribute to reducing the number of demands $\lambda^{d_k}_m$ for contents that need to be forwarded to the DC. Furthermore, using the number of demands $\lambda^{d_k}_m$  and the size of cached contents $d_k$ in collaboration space, we compute bandwidth-saving  through the use of (\ref{eq:caching_reward}). 

Fig. \ref{fig:Bandwidth-saving2} shows the simulation results for bandwidth-saving in terms of Gigabytes (GB). In this figure, from the beginning, bandwidth-saving is nearly zero, and thus MEC server has to cache the contents first. In other words,  MEC caching is based on content prefetching. Therefore, due to the increase in the number of cached contents and demands, the maximum bandwidth-saving of $0.82 \times 10^6$  GB is observed when $a=1.0$. 

Fig. \ref{fig:caching_allocation} shows  the total cache storage utilization in the collaboration space of $12$ MEC servers, where the cache storage utilization depends on the sizes of offloaded data and cache capacity constraints. \textcolor{black}{In Fig. \ref{fig:caching_allocation},  we can see that the cache resources utilization increases with the number of demands until it reaches to $1.13 \times 10^5$ GB (when $a=1.0$). The increase of cache storage utilization results in the increase of cache hits in Fig.  \ref{fig:normalized_cache_hit1} and bandwidth saving in Fig.  \ref{fig:Bandwidth-saving2}.}

\section{Conclusion}
\label{sec:Conclusion}
In this paper, we have proposed a joint communication, computation, caching, and control (4C) framework for big data MEC. In this framework,  MEC servers collaborate to satisfy user demands.  We have formulated the problem as an optimization problem that aims to achieve maximum bandwidth saving while minimizing delay, subject to the local computation capabilities of the user devices, computation deadline, and MEC resource constraints. Therefore, for solving the formulated optimization problem, we have proposed a distributed optimization control algorithm for 4C, which is a modified version of the BSUM method. We have compared the results from distributed optimization control algorithm with the results computed via the Douglas-Rachford splitting method. The simulation results from both methods have shown that our approach increases bandwidth saving and minimizes delay. 

\end{document}